\documentstyle[preprint,aps,eqsecnum,psfig]{revtex}
\newcommand{\SIGN}{\mathop{\rm sgn}\nolimits}
\begin{document}
\tighten
\draft
\preprint{
  \parbox[t]{30mm}{
    hep-th/9812041\\
    DPNU-98-20\\
  }
}
\title{
 Magnetization and dynamically induced finite densities\\
 in three-dimensional Chern-Simons QED}
\author{Taichi Itoh \cite{taichi}}
\address{ 
Department of Physics and Institute of Basic Science, 
Sungkyunkwan University,\\
Suwon 440-746, Korea}
\author{Hiroshi Kato \cite{kato}}
\address{ 
Department of Physics, Nagoya University,
Nagoya 464-8602, Japan}
\date{\today}
\maketitle

\thispagestyle{empty}
\begin{abstract}
In ($2+1$)-dimensional QED with a Chern-Simons term, 
we show that spontaneous magnetization occurs in the context 
of finite density vacua, which are the lowest Landau levels fully 
or half occupied by fermions. 
Charge condensation is shown to appear so as to complement the 
fermion anti-fermion condensate, which breaks the flavor U$(2N)$ 
symmetry and causes fermion mass generation. 
The solutions to the Schwinger-Dyson gap equation show that 
the fermion self-energy contributes to the induction of 
a finite fermion density and/or fermion mass. 
The magnetization can be supported by charge condensation for 
theories with the Chern-Simons coefficient $\kappa=N e^2/2 \pi$, and 
$\kappa=N e^2/4 \pi$, under the Gauss law constraint. 
For $\kappa=N e^2/4 \pi$, both the magnetic field and 
the fermion mass are simultaneously generated
in the half-filled ground state, 
which breaks the U$(2N)$ symmetry as well as the Lorentz symmetry.
\end{abstract}


\setcounter{page}{1}
\narrowtext
\section{Introduction}

\label{sec:intro}
Field theories in ($2+1$)-dimensional space-time
have been intensively studied not only as a laboratory 
for ($3+1$)-dimensional field theory dynamics
but also as effective theories at long distance
in planar condensed matter physics \cite{frad}. 
In particular, some authors have investigated quantum electrodynamics 
in $2+1$ dimensions (QED${}_3$) in connection with the effective field 
theories for high-$T_c$ super-conductivity \cite{dorey}.$\!\!$
\footnote{The application of gauge field theories to 
high-$T_c$ super-conductivity was originally proposed and studied 
in Ref.\ \cite{RVB} in the context of 
the resonating valence bond state. 
The non-Fermi liquid behavior in the normal phase of the 
high-$T_c$ system has been studied with the help of effective 
gauge theories and their renormalization group analysis. 
See Ref.\ \cite{polch} for details.}
In this theory, 
there can be a topological, i.e. metric independent, gauge action, 
known as the Chern-Simons (CS) term, 
which connects a magnetic field $B$ with an electric charge density 
$e\langle\psi^\dagger \psi\rangle$ for the fermion field $\psi$. 
Because of this peculiar property, the CS term is essential 
in the field theoretical understanding of 
the fractional quantum Hall effect, 
which is based on the composite excitation of 
an electron and magnetic fluxes \cite{jain,zhang}. 

A theory whose gauge field action includes both the CS term and the 
Maxwell term was proposed by Deser, Jackiw, and Templeton \cite{deser} 
as an attempt to improve the infrared photon behavior in QED${}_3$.  
In this theory (Chern-Simons QED${}_3$) the coefficient $\kappa$ 
for the CS term gives the photon a gauge invariant mass which 
explicitly violates parity. 
Hosotani \cite{hosotani} showed that spontaneous magnetization 
occurs in Chern-Simons QED${}_3$, because the gauge invariant photon 
mass $\kappa$ is completely screened out 
by parity-violating vacuum polarization effects.
This also means that Lorentz symmetry is broken 
through the induced magnetic field and the resultant massless photon 
plays the role of the Nambu-Goldstone (NG) mode.
In this theory, the Gauss law, 
\begin{equation}
 \kappa B = -e\langle\psi^\dagger\psi\rangle, \label{gauss} 
\end{equation}
follows from the equation of motion. 
Thus the magnetized vacuum corresponds to the ground state with 
finite fermion density, $\langle\psi^\dagger \psi\rangle \neq 0$, 
that is, an occupied lowest Landau level. 

In Ref.\ \cite{hosotani}, however, the finite density vacuum 
configuration was set by hand as a state with a definite filling 
fraction $\nu$ that represents the degrees of occupation of 
fermions in the lowest Landau level. 
The Gauss law (\ref{gauss}) tells us that spontaneous magnetization 
means a phase transition occurs between 
different vacua assigned to specific values of $\nu$. 
However, without any parameter connecting differently occupied vacua, 
it was not clear whether the condensate 
$\langle\psi^\dagger \psi\rangle$, which supports the magnetic field, 
is spontaneously induced or finely tuned. 

In order to clarify the above situation, an external source term 
$\mu\, \psi^\dagger \psi$ was introduced in Ref.\ \cite{IS} 
to control the finite densities in the vacuum 
with a continuous parameter $\mu$. 
From the viewpoint of statistical mechanics, 
$\mu$ plays the role of the chemical potential. 
The parameter $\mu$ has to exceed the fermion mass $m$, 
or the energy of lowest Landau level, 
in order to induce a finite fermion density. 
We notice, however, from the field theoretical viewpoint, 
the chemical potential term $\mu\, \psi^\dagger\psi$ also plays 
the role of a term that explicitly breaks Lorentz symmetry. 
So long as the fermion mass $m$ remains constant, 
we cannot take the Lorentz symmetric limit $\mu \rightarrow 0$ 
while retaining non-zero finite density in the vacuum. 
Alternatively, however, if we take the symmetric limit such that 
$m\le|\mu|\to0$, the finite density vacuum might be spontaneously 
realized through the condensate $\langle\psi^\dagger \psi\rangle$. 
In other words, the finite density vacua are {\it spontaneously} 
realized if and only if the fermion bare mass becomes zero \cite{IS}.

Recently, in Refs.\ \cite{gusynin,klimenko}
it was shown that a strong magnetic 
field enhances the condensate $\langle\bar{\psi}\psi\rangle$, 
or the dynamical generation of fermion mass, 
in ($2+1$)-dimensional four-fermion interaction models. 
Refs.\ \cite{gusynin,gusynin2} explained this effect
as the dimensional reduction
of the phase space for charged particles 
coupled to a strong magnetic field.$\!\!$
\footnote{The enhancement of fermion mass generation 
by a magnetic field was found also in lattice QED$_3$ \cite{FKM} 
and was applied to the effective field theories in planar 
condensed matter systems \cite{semenoff}.}
In a strong magnetic field, the wave function for 
charged fermions is localized within a region whose 
size is given by the magnetic length: $l=1/\sqrt{|eB|}$. 
The fermion in $2+1$ dimensions, therefore,
behaves like that in $0+1$ dimension,
while the photon field is charge neutral and propagates freely 
in $2+1$ dimensional space-time even in the presence of 
the background magnetic field. 
The condensate $\langle\bar{\psi}\psi\rangle$ 
is therefore easily formed much like in 
Bardeen-Cooper-Schrieffer (BCS) theory \cite{gusynin,gusynin2}. 

The Lagrangian of QED${}_3$ with $N$ flavor massless 
four-component fermions has U$(2N)$ symmetry. 
When $N$ is smaller than its critical value $3< N_c < 5$, 
the flavor U$(2N)$ symmetry is spontaneously broken by 
the condensate $\langle\bar{\psi}\psi\rangle$ and the 
fermion acquires a dynamically generated mass \cite{appel,NN}. 
By means of the Schwinger-Dyson (SD) gap equation, 
Shpagin \cite{shpagin} showed that 
a fermion mass is dynamically generated and 
the U$(2N)$ symmetry is spontaneously broken irrespective of $N$ 
in QED${}_3$ with an external magnetic field. 

This result means that the magnetic field catalyzes
the fermion mass generation and 
seems contradictory to the results of Refs.\ \cite{hosotani,IS} 
at first sight, since the magnetic field is spontaneously induced 
only for massless fermions in Refs.\ \cite{hosotani,IS}. 
If both results of Refs.\ \cite{hosotani,IS} and Ref.\ \cite{shpagin} 
are true, the chemical potential $\mu\, \psi^\dagger\psi$ 
should affect the condensate $\langle\bar{\psi}\psi\rangle$ 
in the presence of a background magnetic field. 
It has obviously become an interesting question in Chern-Simons 
QED${}_3$ to study the dynamical generation of both condensates, 
$\langle\bar{\psi}\psi\rangle$ and $\langle\psi^\dagger\psi\rangle$, 
or the fermion mass and the magnetic field, 
as a specific ($2+1$)-dimensional phenomenon. 

In this paper, we will directly show that
the spontaneous magnetization, 
or Lorentz symmetry breaking, occurs as the realization of the 
finite density vacuum in Chern-Simons QED${}_3$. 
We will also examine the possibility that Lorentz symmetry is 
broken in a theory where {\it both massive fermions and a magnetic 
field are spontaneously generated} \cite{IK}. 

The paper is laid out as follows. 
In Section \ref{sec:eff} we derive the effective action
in the large $N$ limit 
that contains the screening of the $N$ flavors of fermions. 
We add to the Lagrangian both the fermion bare mass term 
$m\,\bar{\psi}\psi$ as an explicit breaking source term 
for the flavor U$(2N)$ symmetry, and the chemical potential term 
$\mu\, \psi^\dagger \psi$ for Lorentz symmetry. 
In Section \ref{sec:cond} we directly calculate the condensates, 
$\langle\bar{\psi}\psi\rangle$ and $\langle\psi^\dagger\psi\rangle$, 
as the large $N$ quantum correction of the fermion field 
in the presence of the background magnetic field. 
The effect of the explicit breaking parameters and 
the magnetic field $B$ are taken into account in 
the fermion propagator by using the proper time method
\cite{schwinger}. 
Thus we can see the dependence of $m$ and $\mu$ 
upon the condensates, $\langle\bar{\psi}\psi\rangle$ 
and $\langle\psi^\dagger\psi\rangle$, explicitly. 
It is found from the Gauss law (\ref{gauss}) that the consistency 
condition for nonzero magnetic field places 
a restriction on the CS coefficient $\kappa$ 
in the symmetric limit $\mu\rightarrow 0$, $m\rightarrow 0$. 
In Section \ref{sec:vacum} we calculate the vacuum polarization tensor
in the large $N$ limit in a gauge invariant manner. 
We regularize the parity-violating part, or the induced CS term, 
in a manner consistent with the charge condensation. 
The improved photon propagator including the effect of 
vacuum polarization is also derived.
In Section \ref{sec:sd} we construct and solve
the Schwinger-Dyson (SD) gap equation 
for the fermion self-energy that contains the dynamically induced 
fermion mass $m_d$ and chemical potential $\mu_d$. 
We investigate the solutions to the SD equation in the symmetric limit 
$\mu\rightarrow 0$, $m\rightarrow 0$, 
to clarify whether or not the fermion mass and/or 
the magnetic field are dynamically generated. 
Section \ref{sec:poten} is devoted to the study of vacuum stability. 
We calculate the Cornwall-Jackiw-Tomboulis (CJT) potential 
\cite{corn} for the composite operators, $\bar{\psi}\psi$ and 
$\psi^\dagger \psi$, and the shift of zero-point energies 
for the fermions and the photon. 
In Section \ref{sec:concl} we conclude our paper with a discussion of 
the relation between the results in Refs.\ \cite{hosotani,IS}, 
\cite{shpagin}, and ours.


\section{Large $N$ effective theory in Chern-Simons QED$_3$}

\label{sec:eff}
In this section we construct the large $N$ effective action 
in Chern-Simons QED${}_3$, which includes long range screening 
of charged fermions. 
We use the metric ${\rm diag}(g^{\mu\nu})=(-1,1,1)$ through
out the paper. 

\subsection{The model and the symmetries}
The Lagrangian density of Chern-Simons QED${}_3$ with massless 
fermion is 
\begin{equation}
 {\cal L} = \bar{\psi}\gamma^{\mu}
  \left[ i\partial_\mu +e{\cal A}_\mu \right]\psi
   -\frac{1}{4}{\cal F}_{\mu \nu}{\cal F}^{\mu \nu}
    -\frac{\kappa}{2}\epsilon_{\mu \nu \rho}{\cal A}^{\mu}
     \partial^{\nu}{\cal A}^{\rho}
      -\frac{1}{2\xi}(\partial {\cal A})^2,\label{cslag}
\end{equation}
where the third term is the Chern-Simons term 
which is topological (or metric independent) 
and the last term is a covariant gauge fixing term. 
From the Lagrangian (\ref{cslag}) we obtain the equation of motion
\begin{equation}
 \partial_{\nu} {\cal F}^{\nu \mu}
  -\frac{\kappa}{2}\epsilon^{\mu \nu \rho}{\cal F}_{\nu \rho}
   =-e \bar{\psi}\gamma^{\mu}\psi
\end{equation}
whose vacuum expectation value leads to the Gauss law constraint 
$\kappa B = -e \langle\psi^\dagger\psi\rangle$
[cf.\ Eq.\ (\ref{gauss})], under a given magnetic field. 

The smallest spinor representation of Lorentz group 
in $2+1$ dimensions is provided by two-component spinors. 
A corresponding $2\times2$ matrix representation of the 
Clifford algebra is expressed by the Pauli matrices
\begin{equation}
 \begin{array}{ccc}
  \gamma^0 =\sigma_3, & \gamma^1 =i\sigma_1, & \gamma^2 =i\sigma_2,
 \end{array}
\end{equation}
which obey the anti-commutation relation: 
$\{\gamma^\mu, \gamma^ {\nu}\}=-2g^{\mu\nu}$. 
We have no other $2\times2$ matrix which anti-commutes with all 
$\gamma^\mu$. We, therefore, have no chiral symmetry that 
would be broken by a mass term $m\bar{\psi}\psi$ 
in two-component representation, but the discrete parity symmetry 
is broken by the same mass term. 
To construct a continuous group which resembles the chiral symmetry 
in $3+1$ dimensions, it is necessary to enlarge the spinor 
representation to be the one with four-component spinors. 
Now we have a Clifford algebra given by 
three $4\times4$ matrices, namely,
\begin{equation}
 \gamma^0 = \left(
  \begin{array}{cc}
   \sigma_3 & 0 \\
   0 & -\sigma_3 \\
  \end{array}\right),\,\,
 \gamma^1 = \left(
  \begin{array}{cc}
   i \sigma_1 & 0 \\
   0 & -i \sigma_1 \\
  \end{array}\right),\,\,
 \gamma^2 = \left(
  \begin{array}{cc}
   i \sigma_2 & 0 \\
   0 & -i \sigma_2 \\
  \end{array}\right).
\end{equation}
We can then construct another two $4\times4$ matrices
\begin{equation}
 \gamma^3 = i\left(
  \begin{array}{cc}
     0\; & I \\
     I\; & 0 \\
  \end{array}\right),\,\,
 \gamma^5 = i\left(
  \begin{array}{cc}
     0\; & I \\
    -I\; & 0 \\
  \end{array}\right),
\end{equation}
which anti-commute with $\gamma^0$, $\gamma^1$, and $\gamma^2$. 
Using $\gamma^3$ and $\gamma^5$, we can define a U$(2)$ 
group with generators 
\begin{equation}
 \begin{array}{ccccc}
  \tau_0 = I, & \tau_1 = \gamma^5, & \tau_2 = -i\gamma^3, &
   \mbox{and} & \tau_3 = \gamma^3 \gamma^5.
 \end{array}
\end{equation}
We see that the Lagrangian density for a massless four-component 
fermion is invariant under this global U$(2)$ symmetry. 
The appearance of mass term $m\bar{\psi}\psi$ breaks the U$(2)$ 
symmetry to a subgroup ${\rm U}(1)\otimes{\rm U}(1)$ whose 
generators are $\tau_0$ and $\tau_3$. 
For the system of $N$ flavor four-component fermions, 
global symmetry is extended to U$(2N)$ symmetry 
which is nothing but the direct product of the U$(2)$ symmetry 
and the flavor U$(N)$ symmetry. 
Once again, the mass term $m\bar{\psi}\psi$ breaks this U$(2N)$ 
symmetry to its subgroup ${\rm U}(N)\otimes{\rm U}(N)$. 

In $2+1$ dimensions, the inversion of two space axes 
is equivalent to a $\pi$ rotation on the two-dimensional space 
so that parity transformation should be defined as an inversion 
of one spatial axis: $P,\,(x,y)\to (x,y)_P = (-x,y)$. 
If we define the parity transformation of field operators, 
it should leave the Lagrangian of massless QED${}_3$ remain invariant. 
A specific construction is to transform a two-component spinor 
and the gauge field as follows;
\begin{eqnarray}
 P \psi(t,{\bf x})P^{-1} &=& \sigma_1 \psi(t,{\bf x}_P),\nonumber\\
 P {\cal A}^{\mu}(t,{\bf x})P^{-1} &=& 
  (-1)^{\delta_{\mu 1}}{\cal A}^{\mu}(t,{\bf x}_P).\label{parity}
\end{eqnarray}
A significant feature of the two-component Dirac spinor is 
the fact that mass term $m\bar{\psi}\psi$ changes its sign under 
the parity transformation $P$. As to the gauge field 
the CS term is odd under the above parity operation. 
It has been known that a two-component fermion induces a CS term 
with the same signature as the fermion mass through the one-loop 
vacuum polarization \cite{deser,redlich}.

A convenient method to construct the parity even mass term 
for fermions in $2+1$ dimensions is to use a four-component fermion 
in terms of two two-component fermions $\psi_1$ and $\psi_2$:
\begin{equation}
 \psi=
  \left(
   \begin{array}{c}
    \psi_1 \\ \psi_2 
   \end{array} 
  \right).
\end{equation}
Parity transformation for the four-component spinor $\psi$ 
is defined as 
\begin{equation}
 P\psi(t,{\bf x})P^{-1}=-i\gamma^3 \gamma^1 \psi(t,{\bf x}_P).
  \label{p4}
\end{equation}
It includes the exchange of upper and lower two-component spinors, 
$\psi_1 \rightarrow \psi_2$, $\psi_2 \rightarrow \psi_1$, 
in addition to the $P$ operation for each two-component spinor.
Therefore it is easy to write the parity-conserving mass term: 
\begin{equation}
 m\bar{\psi}\psi=m\psi_{1}^{\dagger}\sigma_3 \psi_{1}
  -m\psi_{2}^{\dagger}\sigma_3 \psi_{2},
\end{equation}
that breaks the U$(2N)$ symmetry to the ${\rm U}(N)\otimes{\rm U}(N)$, 
as well as the parity-violating mass term:
\begin{equation}
 m\bar{\psi}\tau_3 \psi=m\psi_{1}^{\dagger}\sigma_3 \psi_{1}
  +m\psi_{2}^{\dagger}\sigma_3 \psi_{2},
\end{equation}
that is a singlet of the U$(2N)$ group.

According to the theorem found by Vafa and Witten \cite{vafa}, 
it is energetically favorable that the U$(2N)$ symmetry is broken 
to the vector-like global symmetry ${\rm U}(N)\otimes{\rm U}(N)$ 
in the vector-like gauge theories such like QED${}_3$.
On the other hand, the theory of our interest contains the topological 
Chern-Simons term as well as the background magnetic field. 
Since the theory ceases to be vector-like in general, the stability 
of ${\rm U}(N)\otimes{\rm U}(N)$ is not necessarily supported. 
In this paper we constrain ourselves to the case that 
the ${\rm U}(N)\otimes{\rm U}(N)$ symmetry is unbroken, 
therefore the fermion mass is dynamically generated only 
as the parity-conserving mass term and the induced CS term due to the 
fermion mass is completely canceled. 
\subsection{Effective action in the large $N$ limit}
It is known that QED${}_3$ is a super-renormalizable theory and 
its $\beta$-function for a dimensionless coupling constant, 
which is defined as the photon propagation in the large $N$ limit, 
shows a nontrivial infrared fixed point \cite{appel}. 
On the fixed point the gauge coupling $e$ scales 
like $e \sim 1/\sqrt{N}$, which means that low energy effective 
theory is dominated by large $N$ contributions. 
In other words we should not neglect the screening due to 
vacuum polarization at long distance. 
In the following, we construct the effective action in the large 
$N$ limit, which rules infrared dynamics of Chern-Simons QED${}_3$.

According to Ref.\ \cite{appel}, we introduce a dimensionful 
coupling $\alpha = N e^2/4 \pi$ and keep $\alpha$ finite 
when $N$ is taken to infinity so that radiative corrections 
can be added to the effective action successively in $1/N$ expansion. 
What we want to ask is whether or not a magnetic field is 
dynamically induced. 
On the ground state with the induced magnetic field, 
proper excitations are the fermions coupled with the magnetic field 
and the gauge field fluctuation around the magnetic field. 
Taking into account the above expected physics, 
we divide the gauge field ${\cal A}_\mu$ into a 
background field and a fluctuating field; 
${\cal A}_\mu = \sqrt{N}(A^{\rm ext}_\mu + A_\mu)$ with 
$\sqrt{N}A^{\rm ext}_\mu (x)= B x_2 \delta_{\mu 1}$. 
Notice that we have rescaled the gauge field by a factor $\sqrt{N}$, 
which enables us to treat loop expansion as $1/N$ expansion. 
Subsequently, the magnetic field $B$ must scale as $\sqrt{N}$ 
in order that $eB$ remains finite in the large $N$ limit. 

The Lagrangian density (\ref{cslag}) is divided into the gauge part 
and the fermion part; ${\cal L}={\cal L}_{G}+{\cal L}_{F}$,
\begin{eqnarray}
 {\cal L}_{G} &=& -\frac{N}{4}F_{\mu \nu}F^{\mu \nu}
 -\frac{N \kappa}{2}
 \epsilon_{\mu \nu \rho}A^{\mu}\partial^{\nu}A^{\rho}
   -\frac{N}{2\xi}(\partial A)^2 
    -\frac{B^2}{2}-\sqrt{N}\kappa B A^0 ,\nonumber\\
 {\cal L}_{F} &=& \bar{\psi}\left[i\gamma^\mu D_{\mu}[A^{\rm ext}+A]
   -m+\mu\gamma^0\right]\psi,
\end{eqnarray}
where $D_{\mu}[A]:= \partial_{\mu}-ie\sqrt{N}A_{\mu}$ 
is a covariant derivative. 
The propagator for the fermion coupled to $A^{\rm ext}_\mu$ 
is defined by 
\begin{equation}
 S(x,y)= -\left\langle x\left|
  \frac{1}{i\gamma^{\mu} D_{\mu}[A^{\rm ext}]
   -m +\mu \gamma^0}\right|y\right\rangle.\label{fpro0}
\end{equation} 
It is written in the proper time method \cite{schwinger} such as
\begin{equation}
 S(x,y)=\exp
  \left(\frac{ie}{2}\sqrt{N}\,(x-y)^{\mu}A_{\mu}^{\rm ext}(x+y)\right)
    \widetilde{S}(x-y).\label{fpro1}
\end{equation}
Fourier transformation of $\widetilde{S}(x-y)$ is given by 
\cite{dittrich}
\begin{eqnarray}
 \widetilde{S}(k) &=& i\int_{0}^{\infty} d s\,\exp\left[
  -i s \left(m^2 - k_{\epsilon}^2 
   +\frac{\tan (eBs)}{eBs}{\bf k}^2
    \right)\right] \nonumber \\
 && \times \left\{\left[1+\gamma^{1}\gamma^{2}\tan (eBs)\right]
  \left(m + \gamma^{0}k_{\epsilon}\right)
   -(\gamma^{1}k_1 +\gamma^{2}k_2)\sec^{2}(eBs)\right\},\label{fpro2}
\end{eqnarray}
where $k_{\epsilon}:=k^{0}+\mu +i\epsilon\,{\rm sgn}(k^{0})$ 
modifies the $i\epsilon$ prescription to be consistent 
with the shift of Hamiltonian by $\mu$ \cite{chodos}.
In Eq.\ (\ref{fpro2}) the integration over $s$ is the only formal 
expression and, rigorously, we have to choose the sign of $s$ 
according to the sign convention of $k^0 + \mu$ 
so that the integral over $s$ converges. 
The photon propagator from ${\cal L}_{G}$ is 
\begin{equation}
 \Delta^{\mu \nu}(p)=\frac{1}{p^2 + \kappa^2}
  \left[g^{\mu \nu}-\frac{p^\mu p^\nu}{p^2}
   +i\kappa\frac{p_\rho\epsilon^{\mu\nu\rho}}{p^2}
    \right]
     +\xi\frac{p^\mu p^\nu}{(p^2)^2}, 
\end{equation}
which tells us that the CS coefficient $\kappa$ behaves like a 
pole mass of the gauge field. 

The large $N$ effective gauge action is obtained 
by integrating out the fermion field \cite{jackiw}, that is,
\begin{eqnarray}
 \Gamma[A] &=& -i\ln \int [d\psi][d\bar{\psi}]
  \exp\left[\;\int\! d^3 x\; {\cal L}\;\right]\nonumber\\
 &=& \int\! d^3 x\; {\cal L}_{G} -iN\,{\rm Tr\,Ln}
  \Bigl[ -i\gamma^\mu D_{\mu}[A^{\rm ext}+A]+m-\mu\gamma^0 \Bigr]. 
\end{eqnarray}
To expand Tr Ln term with respect to the gauge field $A_\mu$, 
we will extract from it charge condensation, vacuum polarization, 
and nonlocal self-interactions of the gauge field. 
Specifically, it is 
\begin{eqnarray}
 \Gamma[A] &=& -iN\,{\rm Tr\,Ln}\,S^{-1}
  +N\!\int\! d^3 x\;\left[-\frac{B^2}{2N}
   -\frac{\kappa B}{\sqrt{N}}A^0 (x)
    -\frac{1}{2}A^{\mu}(x)\Delta_{\mu\nu}^{-1}(-i\partial_x)A^{\nu}(x)
     \right] \nonumber \\ & &
  +\,N\!\int\! d^3 x\; \left[ 
   -\frac{e\langle\psi^\dagger \psi\rangle}{\sqrt{N}} A^0 (x)
    +\;\frac{1}{2}\int\! d^3 x\; A^\mu (x)
     \Pi_{\mu\nu} (-i\partial_x) A^\nu (x) \right]
      +\widetilde{\Gamma}[A],\label{eff0}
\end{eqnarray}
where $\langle\psi^\dagger \psi\rangle$ is the charge condensation 
induced by a magnetic field and obeys the Gauss law. 
$\widetilde{\Gamma}[A]$ denotes the nonlocal vertices
of the gauge field. 
Vacuum polarization tensor $\Pi_{\mu \nu}$ is given by
\begin{equation}
\Pi_{\mu \nu}(p) = -N e^2 \int\!\! \frac{d^3 k}{i(2 \pi)^3}\,
 {\rm tr}\left[
  \gamma_{\mu}\widetilde{S}(k)\gamma_{\nu}\widetilde{S}(k-p)\right],
   \label{pola1}
\end{equation}
which improves the kinetic term of the gauge field. 

By introducing the improved photon propagator: 
$D_{\mu \nu}(p):= [\Delta^{-1}_{\mu \nu}(p)-\Pi_{\mu \nu}(p)]^{-1}$, 
the above effective action in a complicated form is simplified as
\begin{equation}
 \Gamma[A] = 
  -i N\,{\rm Tr\,Ln}\,S^{-1} -\Omega\,\frac{B^2}{2} 
   -\,\frac{N}{2}\int\! d^3 x\,
    A^{\mu}(x)D_{\mu\nu}^{-1}(-i\partial_x)A^{\nu}(x)
     +\widetilde{\Gamma}[A],\label{eff}
\end{equation}
where $\Omega$ is three-dimensional space-time volume. 
The linear terms for $A_\mu$ in Eq.\ (\ref{eff0}) have been 
canceled out by means of the Gauss law. 
Later in Section \ref{sec:sd} we will study dynamical symmetry 
breaking through the analysis of SD gap equation for the fermion 
self-energy. It can be derived from the effective action (\ref{eff}).
In Section \ref{sec:vacum} vacuum polarization tensor 
$\Pi_{\mu\nu}(p)$ is regularized in a gauge invariant manner, such as
\begin{equation}
 \Pi_{\mu \nu}(p) = (p_\mu p_\nu -p^2 g_{\mu \nu})\,\Pi_{e}(p)
  +\,(p_{\mu}^{\perp} p_{\nu}^{\perp}
   -p^2_{\perp} g_{\mu \nu}^{\perp})\,\Pi_{\perp}(p)
    -i p^{\rho}\epsilon_{\mu \nu \rho}\,\Pi_{o}(p),\label{pola2}
\end{equation}
where $p_{\perp}^{\mu}=(0,p^1,p^2)$ and 
${\rm diag}(g^{\perp}_{\mu\nu})=(0,1,1)$. 
The parity-violating part $\Pi_o (p)$ provides 
a quantum correction to the gauge invariant photon mass $\kappa$ 
in the propagator $D_{\mu\nu}$. 
It will be explained in Section \ref{sec:cond}
that the Gauss law forces the improved photon mass, 
$\kappa-\Pi_o (0)$, to vanish in the presence of a magnetic field. 
Therefore our effective theory in Eq.\ (\ref{eff}) 
has the massless photon, whose propagator is given by $D_{\mu\nu}$, 
as well as the low energy effective theory in QED${}_3$ 
with a magnetic field. 

\section{Condensates and configurations of vacuum}

\label{sec:cond}
In this section we explicitly calculate two condensates, 
$\langle\bar{\psi}\psi\rangle$ and $\langle\psi^\dagger \psi\rangle$, 
for fermions coupled with a magnetic field. 
We can find the possible configurations of vacuum 
(the lowest Landau level) by taking both condensates to 
the symmetric limit $J_\pm:=\mu\pm m\to 0$. 
The condensates are calculated from the fermion propagator 
(\ref{fpro2}) which contains the explicit symmetry breaking 
parameters, $\mu$ and $m$. 
We shall find various patterns of symmetry breaking 
according to different approaches to the symmetric limit $J_\pm\to 0$. 
It is also clarified by applying the result of calculation to 
the Gauss law that possible values of the CS coefficient 
$\kappa$ are restricted to $|\kappa|=0$, $\alpha$, and $2\alpha$. 

The propagator for $N$ flavor fermions $\psi_i$ ($i=1,...,N$) 
coupled with a magnetic field is defined by
\begin{equation}
  S(x,y)\,\delta_{ij} :=
  i\,\langle 0|T\psi_i (x)\bar{\psi}_j (y)|0 \rangle,
  \label{defpro}
\end{equation}
where we assume the ${\rm U}(N)\otimes{\rm U}(N)$ symmetry
and extract the flavor indices as a diagonal metric $\delta_{ij}$. 
$S(x,y)$ in Eq.\ (\ref{defpro}) is equivalent to the definition 
in Eq.\ (\ref{fpro0}). 
We introduce the functions, ${\cal J}$ and ${\cal J}_0$,
which are defined by the following equations:
\begin{eqnarray}
 {\cal J}(m, \mu) &:=& -\lim_{y \to x}{\rm tr}\;
   \langle 0|T\psi_i (x)\bar{\psi}_i (y)|0\rangle \nonumber\\
    &=& iN\;{\rm tr}\; S(x,x),\label{cond:11}\\
 {\cal J}_0 (m, \mu) &:=& -\lim_{y \to x}{\rm tr}\;\gamma^{0}
   \langle 0|T\psi_i (x)\bar{\psi}_i (y)|0\rangle \nonumber\\
    &=& iN\;{\rm tr}\;\gamma^{0} S(x,x).\label{cond:12}
\end{eqnarray}
The condensates, $\langle\bar{\psi}\psi\rangle$ and 
$\langle\psi^\dagger \psi\rangle$, are identified as ${\cal J}$ 
and ${\cal J}_0$ in the symmetric limit $J_\pm\to 0$ such as
\begin{equation}
  \langle\bar{\psi}\psi\rangle =
  \lim_{J_\pm\to\,0}{\cal J} (m,\mu),\;\;\; 
  \langle\psi^\dagger \psi\rangle =
  \lim_{J_\pm\to\,0}{\cal J}_0 (m,\mu).
\end{equation}
If the condensate $\langle\bar{\psi}\psi\rangle$ 
($\langle\psi^\dagger \psi\rangle$) remains nonzero, 
the vacuum is realized as a non-singlet state of U($2N$) symmetry 
(Lorentz symmetry). Precisely, the symmetry is spontaneously broken 
by the condensate though the U($2N$) symmetry (Lorentz symmetry) 
is restored at the tree-level Lagrangian in the symmetric limit 
$J_\pm\to0$ \cite{IS}. 
\subsection{Calculations of condensates}
Since the condensates will be computed in terms of 
the fermion propagator (\ref{fpro2}), 
a comment about the meaning of Wick rotation, $k^0 =i\bar{k}$, 
in the presence of $\mu$ should be noted. 
The improved $i\epsilon$ prescription in Eq.\ (\ref{fpro2}) always 
forces poles of the fermion propagator on a complex $k^0$ plane 
to be located outside the contour which connects the real axis to 
the imaginary axis. 
Therefore the Euclidean propagator becomes
\begin{eqnarray}
 \widetilde{S} (k_E) &=& \int_{0}^{\infty} d s\,\exp\left[
  -s \left(m^2 + (\bar{k}-i\mu)^2 
   +\frac{\tanh \tau}{\tau}k_{\perp}^2
    \right)\right] \nonumber \\
 &&\times \left\{ \left[ 
  1-i\gamma^{1}\gamma^{2}\,{\rm sgn}(eB)\tanh \tau \right]
   \left[ m + i\gamma^{0}(\bar{k}-i\mu) \right]
    -\gamma_{\perp}k_{\perp}\;{\rm sech}^{2} \tau\right\},
     \label{fproE}
\end{eqnarray}
by using $\tau:=|eB|\,s$. 
We will carry out analytic continuation 
$\bar{k}\to\bar{k}-i\mu$ after integrating over a proper time $s$. 
Hereafter the shifted momentum $\bar{k}-i\mu$ in the proper time 
integration necessarily implies the above agreement 
even if we do not indicate it explicitly. 

\begin{figure}
\centerline{\hbox{\psfig{file=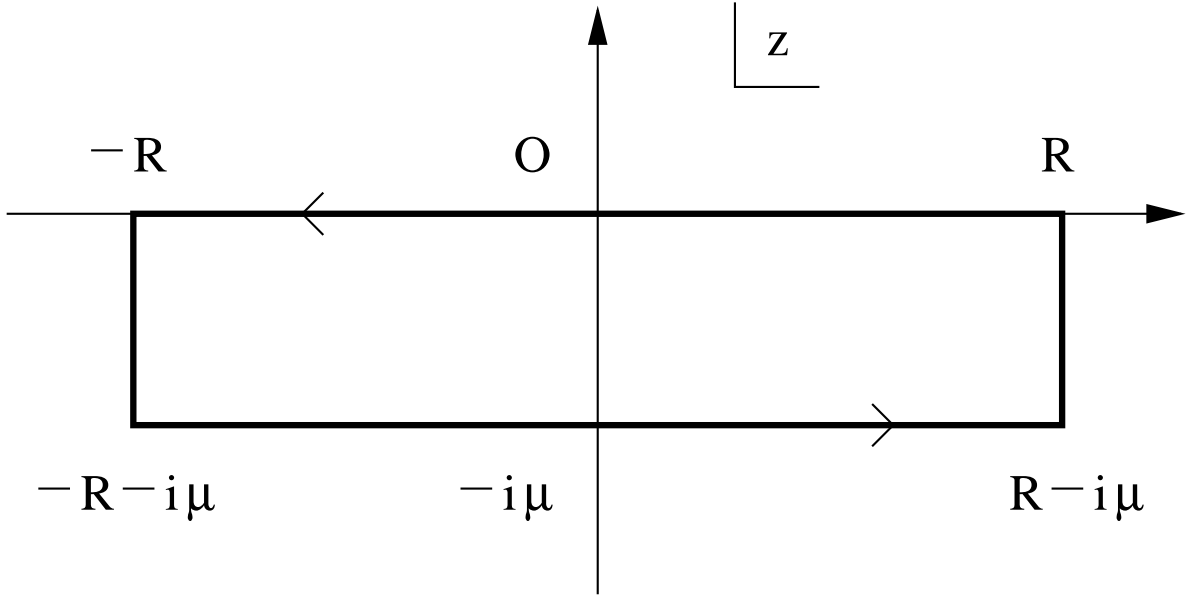,height=8cm}}}\vspace{5mm}
\caption{ The contour of integration, $C_\mu$, 
on the complex $z$-plane (for $\mu > 0$).}\label{cont}
\end{figure}

Now we calculate the condensates ${\cal J}$ and ${\cal J}_0$. 
They are written in terms of the Euclidean propagator 
$\widetilde{S}(k_E)$ as follows;
\begin{eqnarray}
 {\cal J} (m, \mu) &=& -N\int\frac{d^3 k_E}{(2\pi)^3}\;
  {\rm tr}\;\widetilde{S}(k_E) \nonumber\\ 
   &=& -\frac{N}{2\pi^2}|eB|\,m\int_{-\infty}^{\infty}\!d\bar{k}
    \int_{1/\Lambda^2}^{\infty}\!ds\;
     e^{-s\left[(\bar{k}-i\mu)^2 +m^2\right]}\coth \tau,
      \label{cond:21}\\
 {\cal J}_0 (m, \mu) &=& -N\int\frac{d^3 k_E}{(2\pi)^3}\;
  {\rm tr}\;\gamma^{0}\widetilde{S}(k_E) \nonumber\\ 
   &=& -\frac{N}{2\pi^2}|eB|\,i\int_{-\infty}^{\infty}\!d\bar{k}\;
    (\bar{k}-i\mu)\int_{0}^{\infty}\!ds\;
     e^{-s\left[(\bar{k}-i\mu)^2 +m^2\right]}\coth \tau,
      \label{cond:22}
\end{eqnarray}
where the Gaussian integrals for ${\bf k}$ have been done 
and a cut-off scale $\Lambda$ has been introduced to regularize 
the ultraviolet divergence in ${\cal J}(m,\mu)$. 
To extract finite density effects, we first decompose the 
integrands into Landau levels by using an identity: 
\begin{equation}
 \coth \tau \equiv 1 + 2 \sum_{n=1}^{\infty}e^{-2n\tau}.
\end{equation}
Then we integrate out the proper time $s$ and get 
\begin{eqnarray}
 {\cal J} (m, \mu) &=& {\cal J} (m, 0)
  -\frac{N}{2\pi^2}|eB|\,m\left[\,
   \oint_{C_\mu} dz\;
    \frac{1}{z^2 +m^2}+2\sum_{n=1}^{\infty}\;
     \oint_{C_\mu} dz\;
      \frac{1}{z^2 +{\cal E}_{n}^2}\right],\label{cond:31}\\
 {\cal J}_0 (m, \mu) &=& -\frac{N}{2\pi^2}|eB|\left[\;
  \oint_{C_\mu} dz\;
   \frac{iz}{z^2 +m^2}+2\sum_{n=1}^{\infty}\;
    \oint_{C_\mu} dz\;
     \frac{iz}{z^2 +{\cal E}_{n}^2}\right],\label{cond:32}
\end{eqnarray}
where ${\cal E}_{n}:=\sqrt{m^2+2n|eB|}$ is the $n$-th Landau level. 
The above integrations are carried out along a contour 
$C_\mu$ on a complex $z$-plane depicted in Fig.\ \ref{cont}. 
The contribution at zero density appears only in the condensate 
$\langle\bar{\psi}\psi\rangle$ through ${\cal J}(m,0)$
which is given by
\begin{equation}
 {\cal J}(m,0) = -\frac{N}{2\pi^{3/2}}|eB|\,m
  \int_{1/\Lambda^2}^{\infty}\frac{ds}{s^{1/2}}\;
   e^{-s m^2}\coth \tau.
\end{equation}
Using contour integrals shown in Appendix \ref{sec:appenA}, 
we obtain the following results;
\begin{eqnarray}
 {\cal J} (m, \mu) &=& -\frac{N}{\pi^{3/2}}\,\Lambda m
  -\frac{N}{2\pi}|eB|\left[1+\sqrt{2}\;ml\;\zeta
   \left(\frac{1}{2},\frac{(ml)^2}{2}+1\right)\right]
 \nonumber\\ & & 
  +\frac{N}{2\pi}|eB|\left[\theta(|\mu|-m)
   +2\sum_{n=1}^{\infty}\frac{m}{{\cal E}_n}\;
    \theta(|\mu|-{\cal E}_n)\right],\label{cond:41}\\
 {\cal J}_0 (m, \mu) &=& \frac{N}{2\pi}|eB|\;{\rm sgn}(\mu)
  \left[\theta(|\mu|-m)+2\sum_{n=1}^{\infty}
   \theta(|\mu|-{\cal E}_n)\right],\label{cond:42}
\end{eqnarray}
where we have summed up the Landau levels with the 
generalized Riemann zeta function and the value of step function 
$\theta(x)$ at $x=0$ is chosen to be $1/2$ at zero temperature 
limit of Fermi-Dirac distribution. 

We notice that ${\cal J}_0$ reduces to zero 
as we take the limit, $B\to0$ $(l\to\infty)$, in Eq.\ (\ref{cond:42}). 
The charge condensation ${\cal J}_0$ is known to be unaffected 
by higher-order corrections \cite{lykken}. 
Therefore the above one-loop result is exact as we shall make 
detailed comment later on in this section. 
\subsection{Configurations of vacuum}
\label{sec:vacuum-config}
It should be noticed that we can read a filling factor $\nu$ as a 
function of $\mu$ from Eq.\ (\ref{cond:42}) \cite{IS,lykken,zeitlin}. 
Since the wave function for fermions in the lowest Landau level 
spreads out within a size of the magnetic length $l=1/\sqrt{|eB|}$, 
the number of states for an electron is estimated as $1/2\pi l^2$ 
per unit area. 
We also know that $\mu >0$ $(\mu <0)$ induces a finite density of 
fermion (anti-fermion) whose spin indicates the up- (down-) state 
in the lowest Landau level. 
$\langle\psi^\dagger \psi\rangle$ is rewritten as
\begin{equation}
  \langle\psi^\dagger \psi\rangle
  = \frac{N}{2\pi l^2}(\nu_{+}-\nu_{-}), 
 \label{fill}
\end{equation}
where $\nu_{+}$ $(\nu_{-})$ denotes the filling factor of 
fermion (anti-fermion) whose spin indicates the up- (down-) state. 
Comparing Eq.\ (\ref{cond:42}) with Eq.\ (\ref{fill}), we read 
\begin{equation}
 \nu_{\pm}=\theta(\pm\mu)\;\theta(|\mu|-m) \label{fill2}.
\end{equation}

\begin{figure}
\centerline{\hbox{\psfig{file=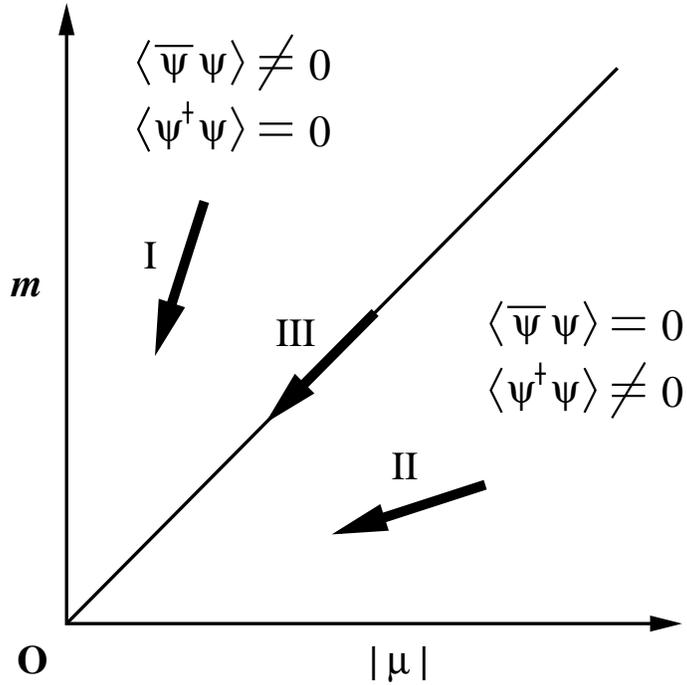,height=9cm}}}\vspace{5mm}
\caption{ Possible approaches to the symmetric limit 
on the ($|\mu|,m$)-plane: I. $|\mu|<m\to0$, II. $m<|\mu|\to0$, and 
III. $|\mu|=m\to0$.} \label{syml}
\end{figure}

Three different approaches to the symmetric limit are possible on 
a $(|\mu|, m)$ plane, namely, $|\mu|<m\to0$, $m<|\mu|\to0$, 
and $|\mu|=m\to0$, in which both condensates,
$\langle\bar{\psi}\psi\rangle$ and $\langle\psi^\dagger \psi\rangle$, 
exhibit the different patterns of symmetry breaking 
(see Fig.\ \ref{syml}) \cite{IS}. 

For $|\mu|<m\to0$, we approach to the origin of the $(|\mu|, m)$ plane 
along a line inside the upper area: $|\mu|<m$. 
The condensates can be read as
\begin{equation}
 \langle\bar{\psi}\psi\rangle = -\frac{N}{2\pi l^2},\;\;\;
  \langle\psi^\dagger \psi\rangle = 0,\label{cond1}
\end{equation}
which shows spontaneous breaking of the U$(2N)$ symmetry on an empty 
vacuum, that stands for the lowest Landau level with $\nu_{\pm}=0$. 
If we believe the Gauss law (\ref{gauss}) naively, the generation 
of magnetic field is available only when charge condensation occurs. 
Since we have no charge condensation seen in Eq.\ (\ref{cond1}), 
we expect no spontaneous magnetization and, subsequently, 
no spontaneous breaking of the Lorentz symmetry.
However, we want to postpone our conclusion until 
we study the vacuum energy shift due to the magnetic field 
in Section \ref{sec:poten}.

For $m<|\mu|\to0$, we choose a line inside the lower area, $m<|\mu|$, 
to approach the symmetric limit. 
We get an opposite pattern of condensates to the former case, namely,
\begin{equation}
 \langle\bar{\psi}\psi\rangle = 0,\;\;\;
  \langle\psi^\dagger \psi\rangle = 
   \frac{N}{2\pi l^2}\;{\rm sgn}(\mu),\label{cond2}
\end{equation}
which mean the vacuum, or the lowest Landau level, has been filled 
up as it is read from $(\nu_+,\nu_-)=(1,0)$, or $(0,1)$, 
and the U$(2N)$ symmetry has been restored.$\!\!$
\footnote{
The restoration of U$(2N)$ symmetry in a given finite density was 
also pointed out in Ref.\ \cite{gusynin} in ($2+1$)-dimensional 
Nambu-Jona-Lasinio model with a magnetic field.}
This time we have obtained a finite density vacuum so that the 
magnetic field is supported by nonzero charge condensation. 
Therefore the magnetic field can be generated dynamically. 
If the gap equation for fermion self-energy favors a finite 
density solution, Lorentz symmetry is spontaneously broken through 
the induced magnetic field. 

For $|\mu|=m\to0$, the symmetric limit is taken along the boundary 
line of $|\mu|=m$ on the $(|\mu|, m)$ plane. 
Both condensates remain finite and are given by
\begin{equation}
 \langle\bar{\psi}\psi\rangle = -\frac{N}{4\pi l^2},\;\;\;
  \langle\psi^\dagger \psi\rangle = 
   \frac{N}{4\pi l^2}\;{\rm sgn}(\mu),\label{cond3}
\end{equation}
which mean that the U$(2N)$ symmetry is broken spontaneously in the 
half-filled vacuum with $(\nu_+,\nu_-)=(1/2,0)$, or $(0,1/2)$. 
As to the Lorentz symmetry, it might also be broken spontaneously 
through the induced finite density which supports the magnetic 
field through the Gauss law. 

We notice that the two condensates always appear 
in such a complementary way as to keep the combination 
$\langle\bar{\psi}\psi\rangle-|\langle\psi^\dagger \psi\rangle|$ 
as a nonzero constant: $-N/2\pi l^2$. 
Moreover the above result depends only on the kinematics of fermions 
in the lowest Landau level.
Even if we have no attractive interaction of photon exchange, 
the magnetic field alone can cause the nonzero condensate: 
$\langle\bar{\psi}\psi\rangle-|\langle\psi^\dagger \psi\rangle|$. 
This is a specific phenomenon realized only in $(2+1)$-dimensions, 
and, for $\langle\bar{\psi}\psi\rangle$, the same discussion was 
already made in Ref.\ \cite{gusynin}.
\subsection{Consistency condition}
In Chern-Simons QED${}_3$, charge condensation 
$\langle\psi^\dagger \psi\rangle$ is necessarily connected with 
a magnetic field through the Gauss law (\ref{gauss}), while 
Eqs.\ (\ref{cond1}-\ref{cond3}) tell us that the magnetic field 
keeps $\langle\psi^\dagger \psi\rangle$ as quantized values 
in a unit of $1/2\pi l^2$ even in the symmetric limit. 
Therefore, the Gauss law places a restriction on the CS coefficient 
$\kappa$ in order to retain the magnetic field in the symmetric limit.
Substituting ${\cal J}_0 (m, \mu)$, given by Eq.\ (\ref{cond:42}), 
for $\langle\psi^\dagger \psi\rangle$ in the Gauss law (\ref{gauss}), 
we obtain in the case of $|\mu| < {\cal E}_{1}$ the equation
\begin{equation}
 B\Bigl[ \kappa +2\alpha\,{\rm sgn}(\mu eB)\,
  \theta(|\mu|-m) \Bigr]=0,
  \label{consis}
\end{equation}
which is regarded as the self-consistency condition for a 
given magnetic field $B$.
In order to keep the nonzero magnetic field in our system, 
$\kappa$ should be restricted to 
\begin{equation}
\kappa=\left\{
\begin{array}{ll}
0 & \mbox{for}\;\;|\mu|<m\to 0 \\ 
-\alpha\,{\rm sgn}(\mu eB) & \mbox{for}\;\;|\mu|=m\to 0 \\
-2\alpha\,{\rm sgn}(\mu eB) & \mbox{for}\;\;m<|\mu|\to 0\\
\end{array}\right..\label{kappa}
\end{equation} 
If we start from the theories with $|\kappa|=0$, $\alpha$, 
and $2\alpha$, which allow the penetration of nonzero magnetic field, 
then each symmetric limit $(|\mu|, m)\to(0,0)$ is 
uniquely determined as $|\mu|<m\to 0$, $|\mu|=m\to 0$, or 
$m<|\mu|\to 0$, respectively. 

The above constraint for $\kappa$ is also interpreted as follows. 
It will be shown in Section \ref{sec:vacum} that, 
except for $|\mu| = {\cal E}_{n}$ ($n=1,2,\dots\,$), 
the charge condensation ${\cal J}_0 (m, \mu)$ is related to the 
parity-violating part of vacuum polarization $\Pi_{o}(p)$ 
through the relation \cite{lykken}: 
\begin{equation}
 -e {\cal J}_0 (m, \mu) = B\, \Pi_{o}(0).
\end{equation}
The constraint (\ref{consis}) is interpreted as 
$B \left[\kappa-\Pi_{o}(0)\right]=0$ which means that 
a magnetic field applied to the system can penetrate it 
if and only if the effective photon mass, 
$\kappa_{\rm eff}:=\kappa-\Pi_{o}(0)$, becomes zero 
\cite{hosotani,hosotani2,NS}. 
Otherwise, the system with $\kappa \neq \Pi_{o}(0)$ excludes 
a magnetic field irrespective of its origin. 
In this case we should set magnetic field zero and investigate 
dynamical generation of the parity-violating fermion mass 
as well as U$(2N)$ symmetry breaking \cite{hong,kondo}.$\!\!$
\footnote{It was pointed out in Ref.\ \cite{kondo} that spontaneous 
breaking of the U$(2N)$ symmetry turns out to be a first-order phase 
transition in the theories with $\kappa \neq \Pi_{o}(0)$.}
In the theories with $\kappa=\Pi_o (0)$, we cannot exclude the 
possibility of spontaneous magnetization since we achieves 
the symmetric limit which allows the penetration of a given 
magnetic field in each theory [cf.\ Eq.\ (\ref{kappa})]. 
We notice that the above statements are not affected by any 
higher-order correction by virtue of the non-renormalization theorem 
of the induced CS coefficient $\Pi_o (0)$ 
(Coleman-Hill theorem \cite{coleman}) even with finite density 
and magnetic field \cite{lykken,SSW}. 

In Section \ref{sec:sd} we investigate dynamical generation of 
fermion mass and magnetic field for theories with 
$|\kappa|=0$, $\alpha$, and $2\alpha$. 
We shall confirm that the fermion mass $m_d$ and the chemical 
potential $\mu_d$ are dynamically generated in the symmetric limit 
$(|\mu|,m)\to(0,0)$. 
In order to retain the consistency with a nonzero magnetic field, 
$m_d$ and $\mu_d$ should satisfy the same patterns 
as the explicit breaking parameters $m$ and $\mu$ do 
in Eq. (\ref{kappa}) for each $\kappa$.

\section{Vacuum polarization}

\label{sec:vacum}
This section is devoted to the gauge invariant derivation 
of the vacuum polarization tensor $\Pi_{\mu \nu}$ defined by 
Eq.\ (\ref{pola1}), that is
\begin{equation}
 \Pi_{\mu \nu}(p) = -N e^2 \int\!\! \frac{d^3 k}{i(2 \pi)^3}\;
  {\rm tr}\left[
   \gamma_{\mu}\widetilde{S}(k)\gamma_{\nu}\widetilde{S}(k-p)\right].
    \label{pola3}
\end{equation}
It should be regularized so as to be gauge invariant such as
\begin{equation}
 \Pi_{\mu \nu}(p) = (p_\mu p_\nu -p^2 g_{\mu \nu})\,\Pi_{e}(p) 
  +\,(p_{\mu}^{\perp} p_{\nu}^{\perp}
   -p^2_{\perp} g_{\mu \nu}^{\perp})\,\Pi_{\perp}(p)
    -i p^{\rho}\epsilon_{\mu \nu \rho}\,\Pi_{o}(p).\label{pola4}
\end{equation}
As mentioned in Section \ref{sec:cond}, effects of an explicit 
breaking parameter $\mu$, or equivalently a chemical potential, 
are included through an analytic continuation, 
$\bar{k}\to\bar{k}-i\mu$, of the third component of an Euclidean 
loop momentum. 
As far as we derive parity-conserving parts of vacuum polarization, 
they are obtained through the above analytic continuation 
from zero density results. 
In the following we divide the vacuum polarization tensor 
into zero density parts and finite density parts, that is,
\begin{eqnarray}
 \Pi_{e}(p) &=& \Pi^{\rm z}_{e}(p)+\Pi^{\rm f}_{e}(p), \nonumber\\
  \Pi_{\perp}(p) &=& \Pi^{\rm z}_{\perp}(p)+\Pi^{\rm f}_{\perp}(p), 
   \label{pola4.5}\\
    \Pi_{o}(p) &\equiv& \Pi^{\rm f}_{o}(p), \nonumber
\end{eqnarray}
and derive them separately.
In our setting of $N$ flavor four-component fermions 
the parity-violating part $\Pi_{o}(p)$ appears only as 
a finite density effect$\!\!$
\footnote{
Since in zero density a parity-violating part necessarily 
appears as an integration of an odd function of $k^0$ 
which is the zeroth component of the loop momentum, 
it vanishes after an integration over $k^0$. 
In our setting of fermions the parity violation due to the fermion 
bare mass $m$ is completely canceled out.}
and is uniquely determined through a gauge invariant constraint 
between $\Pi_{o}(0)$ and charge condensation. 
\subsection{Gauge invariant regularization}
The regularization of $\Pi_{\mu\nu}$ is worked out in the same gauge 
invariant manner as in the case without magnetic field. 
First we set $\mu=0$ and mix two proper times of $\widetilde{S}(k)$ 
and $\widetilde{S}(k-p)$ by means of a parameter integration. 
Then we carry out the Gaussian integrals for spatial components 
of the loop momentum: ${\bf k}$. 
We find that parity-conserving parts appear only as even terms 
for $k^0$ and include ultraviolet divergences, while a 
parity-violating part appears only as a linear term for $k^0$ 
and is finite. 

As to the parity-conserving parts we see that the ultraviolet 
divergences appear as terms with the lower power of 
proper time $s$, namely, $s^{-3/2}$ in $2+1$ dimensions. 
Integrating out the proper time $s$ by part, 
we can subtract the ultraviolet divergences as boundary values 
of the integration over $s$. 
Then we find that remaining finite contributions satisfy 
the transverse form in Eq.\ (\ref{pola4}). 
Zero density parts of $\Pi^{\rm z}_e$ and $\Pi^{\rm z}_\perp$ 
are determined as \cite{dittrich}
\begin{eqnarray}
 \Pi^{\rm z}_e (p) &=& \frac{\alpha}{2\sqrt{\pi}}
  \int_{-1}^{1} dv\int_{0}^{\infty} \frac{ds}{s^{1/2}}\,
   e^{-s\left[m^2 + \phi(p)\right]}\;
    \frac{\tau(\cosh \tau v - v\coth \tau\sinh \tau v)}
     {\sinh \tau},\nonumber\\
 \Pi^{\rm z}_\perp (p) &=& \frac{\alpha}{2\sqrt{\pi}}
  \int_{-1}^{1} dv\int_{0}^{\infty} \frac{ds}{s^{1/2}}\,
   e^{-s\left[m^2 + \phi(p)\right]}\;
    \frac{2\tau(\cosh \tau -\cosh \tau v)}{\sinh^3 \tau}
     -\Pi^{\rm z}_e (p),\label{pola5}
\end{eqnarray}
where we have introduced a rescaled variable $\tau:=s\,l^{-2}$ 
and the function $\phi(p)$ is defined by 
\begin{equation}
 \phi(p):= \frac{1-v^2}{4}
  p_{\parallel}^2 +\frac{\cosh \tau -\cosh \tau v}
   {2\tau\sinh \tau}\,p_{\perp}^2.
\end{equation}
In order to derive the finite density contributions to the 
parity-conserving parts we only have to insert Gaussian integral:
\begin{equation}
  \sqrt{\frac{s}{\pi}}\int_{-\infty}^{\infty}\!d\bar{k}
  \,e^{-s\bar{k}^2}=1
\end{equation}
into the expressions of $\Pi^{\rm z}_e$ and $\Pi^{\rm z}_\perp$ 
in Eq.\ (\ref{pola5}) and to perform analytic continuation 
$\bar{k}\to\bar{k}-i\mu$. 

As to the parity-violating part it is given as a combination of 
SO$(2)$ invariant tensors under a given magnetic field, that is,
\begin{eqnarray}
 \Pi^{\rm odd}_{\mu\nu}(p) &=& \frac{\alpha}{2\pi}
  \int_{-\infty}^{\infty} d\bar{k}\,(\bar{k}-i\mu)
   \int_{-1}^{1} dv\int_{0}^{\infty} ds\,
    e^{-s\left[(\bar{k}-i\mu)^2 +m^2 +\phi(p)\right]} \nonumber\\
 && \times \left[ \frac{2\tau(\cosh \tau \cosh \tau v -1)}
  {\sinh^2 \tau}\,p_{\perp}^\rho \left( g^{\parallel}_{\mu 0}\,
   \epsilon^{\perp}_{\nu\rho}-g^{\parallel}_{\nu 0}\,
    \epsilon^{\perp}_{\mu\rho} \right) \right.\nonumber\\ & &
 \left. +\frac{2\tau v \sinh \tau v}{\sinh \tau}\,
  p^0 \epsilon^{\perp}_{\mu\nu}\right]{\rm sgn}(eB), \label{pola5.5}
\end{eqnarray}
which vanishes for $\mu=0$ after an integration over $\bar{k}$. 
Even if we use an identity 
\begin{equation}
 p^{\rho}\epsilon_{\mu\nu\rho} \equiv 
 p_{\perp}^\rho \left(
   g^{\parallel}_{\mu 0}\,\epsilon^{\perp}_{\nu\rho}
   -g^{\parallel}_{\nu 0}\,\epsilon^{\perp}_{\mu\rho} \right)
    -p^0 \epsilon^{\perp}_{\mu\nu}
\end{equation}
in order to extract the gauge invariant tensor 
$p^{\rho}\epsilon_{\mu\nu\rho}$, 
there remains an ambiguity in $\Pi_o (p)$. 
However, in the presence of the magnetic field, there exists a gauge 
invariant constraint between $\Pi_o (0)$ and charge condensation 
$e{\cal J}_0 (m,\mu)$ \cite{lykken}, namely,
\begin{equation}
 e\;\frac{\partial {\cal J}_0 (\mu,m)}{\partial B}
  =-\Pi_o (0).\label{charge}
\end{equation}
By virtue of Eq.\ (\ref{charge}), $\Pi_o (p)$ is uniquely 
determined from the result of Eq.\ (\ref{cond:42}). 

Through the above procedures, 
the functions $\Pi^{\rm f}_e$, $\Pi^{\rm f}_\perp$,
and $\Pi^{\rm f}_o$ 
are determined as
\begin{eqnarray}
 \Pi^{\rm f}_e (p) &=& \frac{\alpha}{2\pi}\int_{-\infty}^{\infty}
  d\bar{k}\int_{-1}^{1} dv\int_{0}^{\infty} ds\,
   \left\{e^{-s\left[(\bar{k}-i\mu)^2 +m^2 +\phi(p)\right]}
    -e^{-s\left[\bar{k}^2 +m^2 +\phi(p)\right]}\right\}\nonumber \\ 
     & & \times\,\frac{\tau(\cosh \tau v - v\coth \tau\sinh \tau v)}
      {\sinh \tau}, \nonumber\\
 \Pi^{\rm f}_\perp (p) &=& 
  \frac{\alpha}{2\pi}\int_{-\infty}^{\infty} d\bar{k}\int_{-1}^{1}
   dv\int_{0}^{\infty} ds\,\left\{e^{-s\left[(\bar{k}-i\mu)^2 +m^2 
         +\phi(p)\right]} -e^{-s\left[\bar{k}^2 +m^2 +\phi(p)\right]}
   \right\}
     \nonumber \\ & & \times\,\frac{2\tau(\cosh \tau -\cosh \tau v)}
      {\sinh^3 \tau}-\Pi^{\rm f}_e (p),\label{pola7}\\
 \Pi^{\rm f}_o (p) &=& \frac{\alpha}{2\pi}\int_{-\infty}^{\infty}
  d\bar{k}\,i(\bar{k}-i \mu)\int_{-1}^{1} dv\int_{0}^{\infty} ds\,
   e^{-s\left[(\bar{k}-i\mu)^2 +m^2 +\phi(p)\right]}
    \nonumber \\ & & \times\,
     \frac{2\tau(\cosh \tau \cosh \tau v -1)}
      {\sinh^2 \tau}\;{\rm sgn}(eB).\nonumber
\end{eqnarray}
As mentioned in Section \ref{sec:cond}, one cannot do the analytic 
continuation $\bar{k}\to\bar{k}-i\mu$ before the proper time $s$ 
is integrated out, and then the order of integrations should be kept 
in Eq.\ (\ref{pola7}).

When the momentum $p$ vanishes, it is possible to integrate out 
the parameter $v$ and the proper time $s$ before the integration 
over $\bar{k}$. Therefore we can look into how the analytic 
continuation works. 
The integrands in Eq.\ (\ref{pola7}) are decomposed into the 
Landau levels by means of identities:
\begin{eqnarray}
 \frac{\tau(\cosh \tau v - v\coth \tau\sinh \tau v)}{\sinh \tau} &=& 
  \tau\,\sum_{n=0}^{\infty}\Bigl[ 1+(2n+1)v \Bigr]
   e^{-(2n+1+v)\tau}+(v\to-v),\nonumber\\
 \frac{2\tau(\cosh \tau -\cosh \tau v)}{\sinh^3 \tau} &=& 
  4\tau\,\sum_{n=1}^{\infty}\left[n^2\,e^{-2n\tau}
   -n(n+1)\,e^{-(2n+1+v)\tau}\right]+(v\to-v),\label{iden1}\\
 \frac{2\tau(\cosh \tau \cosh \tau v -1)}{\sinh^2 \tau} &=& 
  -2\tau\,\sum_{n=0}^{\infty}\left[2n\,e^{-2n\tau}
   -(2n+1)\,e^{-(2n+1+v)\tau}\right]+(v\to-v).\nonumber
\end{eqnarray}
Inserting the above identities into Eq.\ (\ref{pola5}) 
at zero momentum ($p=0$), 
we acquire the following results for zero density parts 
after we integrate out the parameter $v$ and the proper time $s$;
\begin{eqnarray}
 \Pi^{\rm z}_e (0) &=& -2\alpha l \left[ml + 3\sqrt{2}\;
  \zeta\left(-\frac{1}{2}, \frac{(ml)^2}{2}+1\right)
   -\frac{\sqrt{2}}{2}(ml)^2 \,
    \zeta\left(\frac{1}{2}, \frac{(ml)^2}{2}+1\right)\right],
     \nonumber\\
 \Pi^{\rm z}_\perp (0) &=& 2\alpha l \left[ml + \frac{3\sqrt{2}}{2}\;
  \zeta\left(-\frac{1}{2}, \frac{(ml)^2}{2}+1\right)
   +\frac{\sqrt{2}}{8}(ml)^4 \,
    \zeta\left(\frac{3}{2}, \frac{(ml)^2}{2}+1\right)\right],
     \label{pola6}
\end{eqnarray}
where we have summed up contributions from the higher Landau levels 
by means of the generalized Riemann zeta function.
Substituting Eq.\ (\ref{iden1}) into Eq.\ (\ref{pola7}) 
at zero momentum ($p=0$), we also obtain the following expressions 
for the finite density parts;
\begin{eqnarray}
 \Pi^{\rm f}_e (0) &=& -\frac{\alpha}{\pi} \left\{\;
  \oint_{C_\mu} dz\;l^2 \ln l^2 (z^2 +m^2)
   +2\sum_{n=1}^{\infty}\left[\;
    \oint_{C_\mu} dz\;l^2 \ln l^2 (z^2 +{\cal E}^2)
     +2n\oint_{C_\mu} dz\,\frac{1}{z^2 + {\cal E}_{n}^2}
      \right] \right\},\nonumber\\
 \Pi^{\rm f}_\perp (0) &=& 
  \frac{8\alpha}{\pi}\sum_{n=1}^{\infty}
   \left[-n\oint_{C_\mu} dz\,\frac{1}{z^2 + {\cal E}_{n}^2}
    +\frac{n^2}{l^2} \oint_{C_\mu} dz\,
     \frac{1}{\left(z^2 + {\cal E}_{n}^2\right)^2}
      \right]-\Pi^{\rm f}_e (0) ,\label{pola8}\\
 \Pi^{\rm f}_o (0) &=& \frac{2\alpha}{\pi}\,{\rm sgn}(eB)
  \left\{\; \oint_{C_\mu} dz\,\frac{iz}{z^2 + m^2}
   +2\sum_{n=1}^{\infty}\left[\;
    \oint_{C_\mu} dz\,\frac{iz}{z^2 + {\cal E}_{n}^2}
     -\frac{2n}{l^2}\oint_{C_\mu} dz\,
      \frac{iz}{\left(z^2 + {\cal E}_{n}^2\right)^2}    
       \right]\right\}.\nonumber
\end{eqnarray}
When $|\mu|={\cal E}_n$, poles or branch points on a complex 
$z$-plane are located on the contour $C_\mu$ and force us 
to define the integrations as their Cauchy's principal values. 
Subsequently, we obtain (see also Appendix \ref{sec:appenA})
\begin{eqnarray}
 \Pi^{\rm f}_e (0) &=& -2\alpha l 
  \left\{ l(|\mu|-m)\,\theta(|\mu|-m) +2\sum_{n=1}^{\infty}
   \left[ l(|\mu|-{\cal E}_n) -\frac{n}{l{\cal E}_{n}} \right]
    \theta(|\mu|-{\cal E}_n)\right\},\nonumber\\
 \Pi^{\rm f}_\perp (0) &=& 
  4\alpha l \sum_{n=1}^{\infty}\left\{ 
   \left[ \frac{2n}{l{\cal E}_{n}}-\frac{n^2}{(l{\cal E}_{n})^3} 
    \right]\theta(|\mu|-{\cal E}_n) -\frac{n^2}{l(l{\cal E}_{n})^2}\,
     \delta(|\mu|-{\cal E}_n) \right\}
      -\Pi^{\rm f}_e (0),\label{pola9}\\
 \Pi^{\rm f}_o (0) &=& -2\alpha\,{\rm sgn}(\mu eB)
  \left\{ \theta(|\mu|-m)+2\sum_{n=1}^{\infty}
   \left[ \theta(|\mu|-{\cal E}_n)-\frac{n}{l(l{\cal E}_{n})}\,
    \delta(|\mu|-{\cal E}_n) \right] \right\}.\nonumber
\end{eqnarray}

Now we prove Eq.\ (\ref{charge}) that is the gauge invariant 
constraint between the charge condensation and the parity-violating 
part of vacuum polarization and is essential to determine 
the $\Pi_o (p)$ uniquely. 
We see the following relation is obtained from the definition of 
the fermion propagator [cf. Eq.\ (\ref{fpro0})];
\begin{eqnarray}
 \frac{\partial}{\partial B}S(x,y) &=& \int\!d^3 z\,S(x,z) 
  \sqrt{N}e\,\gamma^{\mu}\frac{\partial A^{\rm ext}_{\mu}(z)}
   {\partial B}S(z,y), \nonumber\\ &=& e\!\int\!d^3 z\,z_2\,
    S(x,z)\gamma^1 S(z,y), \label{delB}
\end{eqnarray}
where we have used the explicit form of the gauge potential 
$A^{\rm ext}_{\mu}(z)$ in the second line. 
Applying the above equation to the definition of charge condensation 
in Eq.\ (\ref{cond:12}), we obtain
\begin{eqnarray}
 e\,\frac{\partial}{\partial B}{\cal J}_0 (m,\mu) &=& 
  iN e^2 \!\!\int\! d^3 z\,z_2\,{\rm tr}\left[ \gamma^0 S(x,z)   
   \gamma^1 S(z,x) \right] \nonumber\\
 &=& \int\!\! d^3 p\;\delta^3 (p) \left( 
  x_2 -i\frac{\partial}{\partial p_2} \right)\Pi^{01}(p) = -\Pi_o (0).
\end{eqnarray}

We see that in Eq.\ (\ref{pola9}) delta function singularities 
appear only in the higher Landau levels, ${\cal E}_n$ ($n\geq1$), 
which have gap energies owing to the magnetic field $B$. 
In order to take the symmetric limit $J_\pm\to0$, it is enough 
to constrain the region of parameters, $J_\pm$, 
to be on the lowest Landau level. 
For $|\mu|<{\cal E}_1$ we notice that Eq.\ (\ref{charge}) reduces to 
\begin{equation}
-e {\cal J}_0 (m, \mu) = B\, \Pi_{o}(0),
\end{equation}
which has been used in Section \ref{sec:cond} 
to derive the consistency condition for the Chern-Simons 
coefficient $\kappa$ in the magnetic field.
\subsection{The improved photon propagator}
In Section \ref{sec:sd} we shall investigate dynamical symmetry 
breaking by means of the Schwinger-Dyson gap equation for the 
fermion self-energy in the lowest Landau level. 
According to the planer property of our model, 
the radiative effect of an improved photon propagator is saturated 
in the infrared region of its momentum, which will be 
shown in the following. 

Recall that the improved photon propagator $D_{\mu\nu}$, which is 
deduced from the effective action $\Gamma[A]$ in Eq.\ (\ref{eff}), 
is given by
\begin{equation}
 D^{-1}_{\mu\nu}(p)=\Delta^{-1}_{\mu\nu}(p)-\Pi_{\mu\nu}(p).
\end{equation}
The tensor form of $D_{\mu\nu}$ becomes
\begin{equation}
  D_{\mu\nu}(p) = 
  D_e (p)\left[g_{\mu\nu}-\frac{p_\mu p_\nu}{p^2}\right]
  +D_\perp (p)\left[g^{\perp}_{\mu\nu}
  -\frac{p^{\perp}_\mu p^{\perp}_\nu}{p_{\perp}^2}\right]
  +D_o (p)\;i\epsilon_{\mu\nu\rho}p^\rho 
  +\xi\,\frac{p_\mu p_\nu}{(p^2)^2},
\end{equation}
with a covariant gauge fixing $\xi$. 
The functions $D_e$, $D_\perp$, and $D_o$ are determined as
\begin{eqnarray}
 D_e (p)\,{\cal G}(p) &=& 1 +\Pi_e (p)
  + \frac{p_{\perp}^2}{p^2}\,\Pi_{\perp}(p), \nonumber\\
  D_\perp (p)\,{\cal G}(p)
  &=& -\frac{p_{\perp}^2}{p^2}\,\Pi_{\perp}(p), 
  \label{gpro1}\\
 D_o (p)\,{\cal G}(p) &=& \frac{\kappa -\Pi_o (p)}{p^2}, 
  \nonumber
\end{eqnarray}
where we have introduced the function ${\cal G}(p)$ which is related 
with a determinant of $D^{-1}$ and is defined by
\begin{eqnarray}
 {\cal G}(p) &:=& [1+\Pi_e (p)]
  \left\{p^2 [1+\Pi_e (p)]+p_{\perp}^2 \Pi_\perp (p) 
   \right\} \nonumber\\ & & 
 +\,[\kappa -\Pi_o (p)]^2 \nonumber\\ &\equiv&
  -\xi\,(p^2)^{-2}\,{\rm det}\,D^{-1}(p). \label{det}
\end{eqnarray}
The consistency condition, $\kappa=\Pi_o (0)$, means that 
the function ${\cal G}(p)$ reduces to zero as $p^\mu \to0$ 
in the limit $J_\pm\to0$, where symmetries are restored 
in the tree-level Lagrangian. 
Therefore a photon becomes a massless propagating mode 
which is regarded as a Nambu-Goldstone (NG) mode due to 
the spontaneous breaking of Lorentz symmetry \cite{hosotani}. 

Under a given magnetic field, we have shown that 
there exist three possible theories with the quantized 
CS coefficient $|\kappa|=0$, $\alpha$, and $2\alpha$, 
each of which has a massless photon. 
As far as we are concerned with the infrared behavior of the photon, 
it is enough to approximate the vacuum polarization functions 
with its values at zero momentum. 
From Eq.\ (\ref{pola6}) and Eq.\ (\ref{pola9}) 
the parity-conserving parts $\Pi_e$ and $\Pi_\perp$ are read as
\begin{eqnarray}
 \Pi_e (0) &=& -6\sqrt{2}\,\zeta(-1/2)\,\alpha l, \nonumber\\
 \Pi_\perp (0) &=& 3\sqrt{2}\,\zeta(-1/2)\,\alpha l, \label{pola10}
\end{eqnarray}
which are irrespective of approaches to the symmetric limit 
$J_\pm\to0$. The improved photon propagator behaves, 
for all of the theories with 
$|\kappa|=0$, $\alpha$, and $2\alpha$, like
\begin{eqnarray}
 D_{\mu\nu}(p) &\simeq& \frac{1}{p^2 [1+c\,\alpha l]}
  \left[g_{\mu\nu}-\frac{p_\mu p_\nu}{p^2}\right]
   +\xi\,\frac{p_\mu p_\nu}{(p^2)^2}\nonumber\\ & & 
 +\,\frac{c\,[\,p_{\perp}^2 g^{\perp}_{\mu\nu}
  -p^{\perp}_\mu p^{\perp}_\nu ]}{p^2 [1+c\,\alpha l]
   \left\{2\,p^2 [1+c\,\alpha l]
    -c\,p_{\perp}^2\right\}},\label{impro}
\end{eqnarray}
with a positive constant $c=-6\sqrt{2}\,\zeta(-1/2)\simeq 1.76397$.
\subsection{Expansions in infrared and ultraviolet regions of momentum}
In Section \ref{sec:poten} we shall investigate stability of 
vacua which are realized as possible solutions of SD gap equation, 
through the calculation of vacuum energy. 
For this purpose we need to know the asymptotic behavior 
of vacuum polarization functions both in infrared and 
ultraviolet regions of momentum. 

First we separate the functions into two parts according to 
$p^2 < l^{-2}$ or $p^2 > l^{-2}$ such as
\begin{equation}
 \Pi_i (p) = \Pi^{<}_{i}(p)\;\theta (1-(lp)^2)
  +\Pi^{>}_{i}(p)\;\theta ((lp)^2-1),\label{pola11}
\end{equation}
where $i=e$, $\perp$, and $o$.
The functions $\Pi^{<}_{i}(p)$ are estimated by expanding them 
with respect to $(lp)^2$. In rigorous sense the expansion around 
$p^2 =0$ is possible only if the functions are analytic 
around $p^2 =0$. 
Indeed at zero momentum the functions $\Pi^{<}_{i}(p)$ show 
singularities of delta functions when $|\mu|={\cal E}_n$ 
($n\ge1$) as read from Eq.\ (\ref{pola9}). 
However, in our present study, since the parameter $\mu$ is nothing 
but the probe to pick up the spontaneous breaking of Lorentz 
symmetry and finally we take the limit $J_\pm\to0$, 
we only have to consider the case of $|\mu|<{\cal E}_1$, which is 
free from the delta function singularities at the zero momentum. 
It is also the reason why we believe $\Pi_o (0)$ is unaffected 
by higher-order corrections even when $|\mu|=m$.
In the higher momentum $p^2 > l^{-2}$, the functions $\Pi^{>}_{i}(p)$ 
are expanded with respect to $(lp)^{-2}$ 
by means of the asymptotic expansion. 
Since the magnetic field can be treated as a weak field 
in comparison with the momentum, the expansion with $(lp)^{-2}$ 
matches the one with respect to the magnetic field. 

Now we shall provide a formal expansion for each case. 
In the lower momentum ({\it strong field regime}) we can formally 
expand exponential factors in Eq.\ (\ref{pola5}) and 
Eq.\ (\ref{pola7}) with respect to $(lp)^2$. 
By using $x:=(lp)^2$, they are given by
\begin{eqnarray}
 \Pi^{<}_{e} (p) &=& \alpha l\left[
  L_{e}^{(0)}
   +L_{e}^{(1)}\,x 
     +\,L_{e}^{(2)}\,x^2 
      +\cdots \right], \nonumber\\
 \Pi^{<}_{\perp} (p) &=& \alpha l\left[
  L_{\perp}^{(0)}
   +L_{\perp}^{(1)}\,x 
     +\,L_{\perp}^{(2)}\,x^2 
      +\cdots \right], \label{pola12}\\
 \Pi^{<}_{o} (p) &=& \alpha \left[
  L_{o}^{(0)}
   +L_{o}^{(1)}\,x 
     +\,L_{o}^{(2)}\,x^2 
      +\cdots \right]. \nonumber
\end{eqnarray}
All coefficients $L_{e}^{(n)}$, $L_{\perp}^{(n)}$, 
and $L_{o}^{(n)}$ ($n=0,1,\cdots$) 
are dimensionless and show the delta function singularities 
in $|\mu|={\cal E}_n$ ($n\ge1$), which means breakdown 
of analyticity around $p^2 =0$. 
However, as far as we consider the symmetric limit $J_\pm\to0$, 
such singularities do not appear in any coefficient.
Actually, our vacuum is realized in the symmetric limit as 
the lowest Landau level which is free from the delta function 
singularities.
The first order coefficients are given by
\begin{equation}
 L_{e}^{(0)} = c, \quad L_{\perp}^{(0)} = -\frac{c}{2}, \quad
 \mbox{and}\quad L_{o}^{(0)} = \bar{c}, \label{IRc}
\end{equation}
where, for convenience sake, we have introduced 
$\bar{c}:=\kappa/\alpha$ which becomes $\bar{c}=0,\pm 1$, 
and $\pm 2$ in the symmetric limit.

In the higher momentum ({\it weak field regime}) the functions 
$\Pi^{>}_{i}(p)$ are systematically expanded as a power 
series of $1/x$ as shown in Appendix \ref{sec:appenB}. 
They are formally given as an asymptotic series 
in the symmetric limit such as
\begin{eqnarray}
 \Pi^{>}_{e} (p) &=& \frac{\alpha l}{\sqrt{x}}\left[
  H_{e}^{(0)}
   +H_{e}^{(1)}\,\frac{1}{x}
    +\,H_{e}^{(2)}\,\frac{1}{x^2}
     +\cdots \right], \nonumber\\
 \Pi^{>}_{\perp} (p) &=& \frac{\alpha l}{\sqrt{x}}\left[
  H_{\perp}^{(0)}
   +H_{\perp}^{(1)}\,\frac{1}{x}
    +\,H_{\perp}^{(2)}\,\frac{1}{x^2}
     +\cdots \right], \label{pola13}\\
 \Pi^{>}_{o} (p) &=& \alpha \left[
  H_{o}^{(0)}
   +H_{o}^{(1)}\,\frac{1}{x}
    +\,H_{o}^{(2)}\,\frac{1}{x^2} 
     +\cdots \right]. \nonumber
\end{eqnarray}
According to the even or odd character under the parity, we find 
that coefficients $H_{e}^{(2n+1)}$, $H_{\perp}^{(2n+1)}$, 
and $H_{o}^{(2n)}$ vanish for all nonnegative integers $n$. 
We approximate the vacuum polarization functions as their 
asymptotic expansions up to $O(x^{-1})$ and the coefficients 
are given by (see Appendix \ref{sec:appenB})$\!\!$
\footnote{We use $p$ as $p\equiv\sqrt{p^2}$ except for 
the usage of $p$ as an argument of functions.}
\begin{eqnarray}
 H_{e}^{(0)} &=& \lim_{J_\pm\to\,0} \Bigl[ \theta(m-|\mu|)I(p;m) 
  +\,\theta(|\mu|-m) 
 \,\theta\!\left[ p^2 -4(\mu^2 -m^2) \right]\!
  I(p;\mu) \Bigr], \nonumber\\
 H_{o}^{(0)} &=& \lim_{J_\pm\to\,0}{\rm sgn}(\kappa)\,
  \theta(|\mu|-m)\,\theta\!\left[ p^2 -4(\mu^2 -m^2) \right] 
 \frac{4\left[ p^2 -2(\mu^2 -m^2) \right]}
  {p\sqrt{p^2 -4(\mu^2 -m^2)}}, \label{UVC:1}\\
 H_{\perp}^{(0)} &\equiv& 0,\nonumber
\end{eqnarray}
where the function $I(p;u)$ is given by Eq.\ (\ref{b:15}) 
in Appendix \ref{sec:appenB} and we use 
${\rm sgn}(\kappa)\equiv-{\rm sgn}(\mu eB)$, which obeys 
the Gauss law constraint [cf. Eq.\ (\ref{kappa})].

We notice that $H_{e}^{(0)}$ coincides with the coefficient 
calculated in the absence of the magnetic field. 
This expansion in the weak field regime turns out to match 
the one around $B=0$ (or $l\to\infty$). 
The coefficients are determined in the symmetric limit 
such as (see Appendices \ref{sec:appenB} and 
\ref{sec:appenC} for details)
\begin{equation}
 H_{e}^{(0)} = \frac{\pi}{2}, \quad H_{\perp}^{(0)} = 0, \quad
 \mbox{and}\quad H_{o}^{(1)} = 2\bar{c}. \label{UVC2} 
\end{equation}
We will show in Section \ref{sec:poten}
that the coefficient $H_{o}^{(1)}$, 
which is proportional to CS coefficient $\kappa$, 
essentially contributes to a linear term of the magnetic field $B$ 
in the shift of photon zero-point energy.

\section{Schwinger-Dyson gap equation}

\label{sec:sd}
In this section we study the Schwinger-Dyson gap equation 
for the fermion self-energy and show that dynamical breaking of 
both U$(2N)$ and Lorentz symmetries are realized as one of 
nontrivial solutions to the SD equation. 
As confirmed in Section \ref{sec:vacuum-config}, both condensates 
corresponding to the spontaneous breaking of U$(2N)$ and Lorentz 
symmetries necessarily appear in such a complementary manner
as to keep 
$\langle\bar{\psi}\psi\rangle-|\langle\psi^\dagger \psi\rangle|$ 
at a nonzero constant. 
It is a specific feature of $(2+1)$-dimensional physics due to 
the kinematics of fermions in the magnetic field and this feature 
is also reflected in the structure of the SD equation. 

The fermion self-energy contains, as its dynamical variables, 
a scalar component $m_d$ and a $\gamma^0$ component $\mu_d$, 
which correspond to the dynamically induced fermion mass and 
chemical potential, respectively. 
These variables, $m_d$ and $\mu_d$, are determined as solutions 
to the SD gap equation in the same self-consistent manner as 
the Hartree-Fock equation for energy gap in BCS theory.
The analysis of condensates in Section \ref{sec:vacuum-config}
implies that it is convenient to use the proper combination 
$J_\pm =\mu\pm m$ instead of the explicit breaking parameters, $\mu$ 
and $m$, in order to manifest the spontaneous breaking of symmetries. 
Subsequently, we are led to combinations; 
$\omega_\pm :=(\mu_d \pm m_d)l$ \cite{IK}. 
This seems quite natural assignment once we recognize that 
$-\omega_{-}$ ($\omega_{+}$) corresponds to the energy gap of 
fermions on the lowest Landau level
relative to the chemical potential 
(or Fermi energy) when $\mu_d >0$ ($\mu_d <0$). 
In fact the coupled SD equations for $m_d$ and $\mu_d$ can be 
reduced to two decoupled equations. 
One is the gap equation for $\omega_{+}$ and the other is 
that for $\omega_{-}$ as shown later. 
This is another specific feature of ($2+1$)-dimensional physics 
as that in condensates. 
\subsection{Schwinger-Dyson equation in the ladder approximation}
We start our argument with construction of the SD equation 
in the ladder (bare vertex) approximation. 
In Section \ref{sec:eff} we have derived
the $1/N$ leading effective theory in which 
a magnetic field couples to fermions as a background field. 
Our aim of gap equation analysis is to confirm realization of 
spontaneously magnetized vacuum supported by the condensate 
$\langle\psi^\dagger \psi\rangle\neq0$. 
As in BCS theory, what sorts of excitations appear in a given system 
depends on the ground state which should be realized as one of 
solutions to the gap equation for self-energies of the excitations. 

In our present study the proper excitation is nothing but 
the fermion coupled with a magnetic field. 
Therefore we must construct the SD equation based on the bare fermion 
propagator $S$ defined by Eq.\ (\ref{fpro0})
which should be recognized 
as the one for the proper excitation on the magnetized vacuum. 
As to the photon propagator, it has already been derived from the 
$1/N$ leading effective action (\ref{eff}) and has shown the massless 
photon behavior through the Gauss law as given by Eq.\ (\ref{impro}). 
Thus the SD equation has the same contents as the one in QED${}_3$ 
with a magnetic field \cite{shpagin} apart from the fact that 
the fermion self-energy contains a $\gamma^0$ component 
other than a scalar component. 

\begin{figure}
\centerline{\hbox{\psfig{file=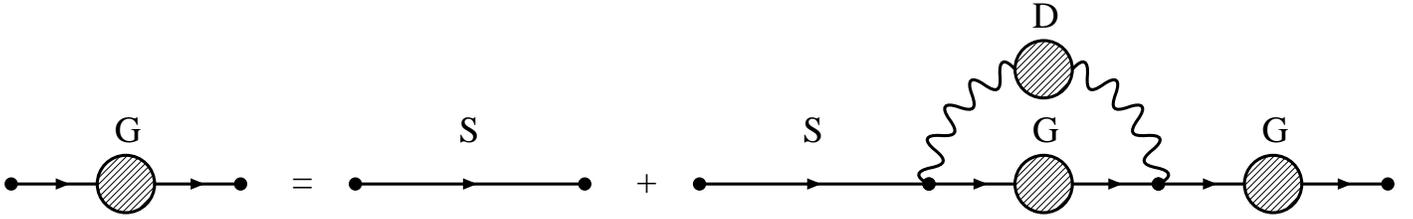,height=3cm}}}\vspace{5mm}
\caption{ The Schwinger-Dyson equation. 
The lines with blob mean the full propagators.} \label{sdeqn}
\end{figure}

The SD equation is given by the recurrent form of
\begin{equation}
  G(x,y) = S(x,y) - ie^2\int d^3 z\,d^3 t\, 
  S(x,z) \gamma^{\mu}G(z,t)\gamma^{\nu}G(t,y)
  D_{\mu\nu}(z-t),\label{3Dsd}
\end{equation}
(see Fig.\ \ref{sdeqn}) where $G$ denotes a full fermion propagator 
which should be consistently determined through the SD equation. 
We assume that the full propagator $G$ also has the same form:
\begin{equation}
G(x,y)=\exp\left(\frac{ie}{2}\sqrt{N}\,(x-y)^{\mu}A^{\rm ext}_{\mu}
(x+y)\right)\widetilde{G}(x-y) \label{fpro3}
\end{equation}
as $S$ in Eq.\ (\ref{fpro1}). 
We substitute the propagators 
into Eq.\ (\ref{3Dsd}) and perform the Fourier transform. 
After integrating out some of spatial coordinates, we obtain
\begin{eqnarray}
  \widetilde{G}(p) &=& \widetilde{S}(p) -\frac{ie^2}{(2\pi)^5}
  \int d^3 k\, d^2 q_{\perp}\, d^2 R_{\perp}
  e^{-iR_{\perp}q_{\perp}}
  \nonumber\\
  &&
  \times 
  \widetilde{S}\left(
    p^0, p_{\perp}+\frac{q_{\perp}}{2} + \frac{eB\bar{R}_\perp}{2}
  \right)
  \gamma^{\mu}
  \widetilde{G}(k)
  \gamma^{\nu}
  \widetilde{G}\left(
    p^0, p_{\perp}-\frac{q_{\perp}}{2} + \frac{eB\bar{R}_\perp}{2}
  \right)
  \nonumber\\
  &&
  \times 
  D_{\mu\nu}\left(
    p^0-k^0, p_{\perp}-k_{\perp}+eB\bar{R}_\perp
  \right),
  \label{sdmom}
\end{eqnarray}
where $R_{\perp}^\mu:=(0,R_1,R_2)$ and 
$\bar{R}_{\perp}^\mu:=(0,-R_2, R_1)$. 

Following Ref.\ \cite{shpagin}, 
we assume the strong magnetic field $m$ ($m_d$) $\ll l^{-1}$, 
so that Eq.\ (\ref{sdmom}) can be simplified owing to decoupling 
of the higher Landau levels ${\cal E}_n (n \ge 1)$. 
The bare fermion propagator $\widetilde{S}(k)$ in Eq.\ (\ref{fpro2}) 
can be decomposed into the Landau level poles \cite{chodos};
\begin{equation}
  \widetilde{S}(k) = \exp\left(-\frac{k_{\perp}^2}{|eB|}\right)
  \sum_{n=0}^{\infty}(-1)^n
  \frac{D_n(eB,k)}{m^2 + 2|eB|n - k_{\epsilon}^2}.
  \label{decomp}
\end{equation}
The function $D_n$ is given by
\begin{eqnarray}
  D_n(eB,k) &=& (m + k_{\epsilon}\gamma^0)
  \left[
  L_n^0 \left(2\frac{k_{\perp}^2}{|eB|}\right)2\Lambda_{+}
  -L_{n-1}^0 \left(2\frac{k_{\perp}^2}{|eB|}\right)2\Lambda_{-}
  \right] \nonumber\\
  && +\; 4k_{\perp}\gamma_{\perp}
  L_{n-1}^1 \left(2\frac{k_{\perp}^2}{|eB|}\right),
\end{eqnarray}
where $L_n^{\alpha}(x)$ are the generalized Laguerre polynomials 
and $\Lambda_{\pm}$ denote the projection operators to spin states:
\begin{equation}
\Lambda_{\pm}:= \frac{1 \mp i\gamma^1\gamma^2\SIGN(eB)}{2}.
\end{equation} 

We can easily see in Eq.\ (\ref{decomp}) that 
under the strong magnetic field, $m \ll l^{-1}$, the lowest Landau level 
dominates and all other higher Landau levels are negligible. 
Therefore, in SD equation (\ref{3Dsd}), we approximate 
the Fourier transform of $\widetilde{S}$ 
with their lowest Landau level contributions, that is 
\begin{equation}
 \widetilde{S}(k) \simeq e^{-l^2 k_{\perp}^2}
 \frac{1}{m-\gamma^0\, k_{\epsilon}}\,2\Lambda_{+}.
\end{equation}
We notice that fermions on the lowest Landau level 
essentially behave like ($0+1$)-dimensional objects. 
It is natural to write down the full fermion propagator 
also in the following ($0+1$)-dimensional form:
\begin{equation}
  \widetilde{G}(k) \simeq e^{-l^2 k_{\perp}^2}\,\widetilde{g}(k^0)
  \,2\Lambda_{+},
\end{equation}
where $\widetilde{g}(k^0)$ is a matrix commutative with $\Lambda_{+}$. 

Substituting $\widetilde{S}$ and $\widetilde{G}$ 
into Eq.\ (\ref{sdmom}),
we carry out $q_{\perp}, R_{\perp}$ integration so that
Eq.\ (\ref{sdmom}) becomes simplified ($0+1$)-dimensional form 
\cite{shpagin}:
\begin{equation}
  \widetilde{g}^{-1}(p^0) = m - \gamma^0(p^0+\mu)
  -\frac{ie^2}{(2\pi)^3}\int_{-\infty}^{\infty} dk^0 \gamma^0 
  \widetilde{g}(k^0)\gamma^0 \widetilde{D}(p^0 -k^0),
  \label{1Dsd}
\end{equation}
where the function $\widetilde{D}$ is defined by
\begin{equation}
 \widetilde{D}(p^0):= -\int d^2 p_\perp 
 e^{-l^2 p_{\perp}^{2}/2}D_{00}(p^0, p_\perp).
 \label{1Dimpro}
\end{equation}
Note that since we assume there is a magnetic field, the Gauss law 
(\ref{gauss}) forces the effective CS coefficient at zero momentum 
to be zero. 
The photon in the magnetic field, therefore, behaves as a 
massless mode in infrared region no matter how we take 
the symmetric limit $J_\pm
\rightarrow 0$. 
According to Eq.\ (\ref{impro}), 
the one dimensional photon propagator $\widetilde{D}$ is
represented by the integral exponential function
which has a logarithmic
behavior in the infrared region of momentum such as  
\begin{equation}
 \frac{e^2 l}{(2\pi)^3}\widetilde{D}(p)\approx 
 -\frac{\alpha_0}{\pi} \ln |lp|,
 \label{infraphoton}
\end{equation}
where $\alpha_0 := \alpha l/N(1+c\,\alpha l)$ and $p:=-ip^0$. 
The constant $c$ denotes parity even vacuum polarization effect, 
$\Pi_e (0)\equiv c\,\alpha l$, and is given by 
$c=-6\sqrt{2}\,\zeta(-1/2)\simeq 1.76397$. 
Thus only the infrared momentum region is relevant
in Eq.\ (\ref{1Dimpro}).
\subsection{Rearrangement of dynamical variables}
The matrix function $\widetilde{g}$ is written in an SO$(2)$ 
invariant form;
\begin{equation}
 \widetilde{g}^{-1}(p^0) = B(p^0) +\widehat{A}(p^0)\,p^0
 -\gamma^0 \left[\widehat{B}(p^0) +A(p^0)\,p^0 
 +i \epsilon\,\SIGN(p^0)\right],
 \label{fng}
\end{equation}
where the functions $A$, $B$, $\widehat{A}$, and $\widehat{B}$ are 
even functions of $p^0$. 
The self-energy part is divided into scalar and $\gamma^0$ 
components as the functions $B$ and $\widehat{B}$, respectively. 
Since $p^0$ is SO$(2)$ invariant in itself, $\widehat{A}(p^0)\,p^0$ 
should be involved in $\widetilde{g}^{-1}$ as an odd function part 
of the scalar component. 
We can set the functions $A(p^0)$ and $B(p^0)$ 
to be positive definite without loss of generality. 
Following the physical implication mentioned in the beginning, 
we rearrange the functions $A$, $B$, $\widehat{A}$, 
and $\widehat{B}$ as
\begin{eqnarray}
 A_\pm (p^0) &:=& A(p^0)\pm \widehat{A}(p^0), \nonumber\\
 B_\pm (p^0) &:=& \widehat{B}(p^0)\pm B(p^0).
\end{eqnarray}
If we recognize that the functions $\pm B_{\pm}/A_{\pm}$
are identified 
with the relative energy of the proper excitation to the Fermi energy, 
it seems natural to assume that $B_{\pm}/A_{\pm}$ have a definite sign 
irrespective of its argument. 
Then we can perform the Wick rotation $k^0=ik$ in the SD equation
without any ambiguity.$\!\!$
\footnote{We use $k$, $p$ instead of $\bar{k}$, $\bar{p}$ only in this 
section to simplify the mathematical formulas.}
Actually $\widetilde{g}(k^0)$ is decomposed into two propagators 
which correspond to the proper excitations in the lowest Landau level;
\begin{eqnarray}
 \widetilde{g}(k^0) &=& 
  \frac{1-\gamma^0}{2}\left[ \frac{1}{B_{+}(k^0) 
   + A_{+}(k^0)\,k^0 + i\epsilon\SIGN(k^0)} \right] \nonumber\\
 && -\, \frac{1+\gamma^0}{2}\left[ \frac{1}{B_{-}(k^0) 
  + A_{-}(k^0)\,k^0 + i\epsilon\SIGN(k^0)} \right],\label{SDpoles}
\end{eqnarray}
whose poles are located on a complex $k^0$ plane at
\begin{equation}
 k^0 = \left\{
 \begin{array}{rcl}
  -|B_{\pm}/A_{\pm}|+i\epsilon & \rm{for} & B_{\pm}/A_{\pm} > 0 \\
   |B_{\pm}/A_{\pm}|-i\epsilon & \rm{for} & B_{\pm}/A_{\pm} < 0
 \end{array}\right..
\end{equation}
Thus we can perform the Wick rotation without any residual 
contributions from poles.

Substituting Eq.\ (\ref{fng}) into the SD equation (\ref{1Dsd}) 
and performing the Wick rotation $k^0 =ik$, $p^0 =ip$, 
we obtain two sets of coupled integral equations \cite{IK}
\begin{eqnarray}
 B_{\pm}(p) -J_\pm &=& \frac{e^2}{(2\pi)^3}
  \int_{-\infty}^{\infty} dk
   \frac{B_\pm(k)}{A_{\pm}^{2}(k)\,k^{2}
    +B_{\pm}^{2}(k)}
     \widetilde{D}(p-k),\label{PreGapEq1}\\
 p\left[1-A_{\pm}(p)\right] &=& \frac{e^2}{(2\pi)^3}
  \int_{-\infty}^{\infty} dk
   \frac{A_\pm(k)\,k}{A_{\pm}^{2}(k)\,k^{2}
    +B_{\pm}^{2}(k)}
     \widetilde{D}(p-k),\label{PreGapEq2}  
\end{eqnarray}
which are divided into the coupled equations for $(A_{+},B_{+})$ 
and those for $(A_{-},B_{-})$ by virtue of the specific feature 
of ($2+1$)-dimensions [cf.\ Eq.\ (\ref{SDpoles})]. 
It is obvious that the above equations have trivial solutions 
$B_{\pm}(p)\equiv0$ in the symmetric limit $J_\pm \to0$. 

Firstly we determine $A_{\pm}(p)$ for the trivial solutions, 
$B_{\pm}(p)\equiv0$, as well as for the nontrivial solutions. 
We differentiate both sides of Eq.\ (\ref{PreGapEq2}) 
with respect to $p$ and obtain
\begin{equation}
 1-A_{\pm}(p)-p\,A_{\pm}^{\prime}(p)=
  -\frac{\alpha_0}{\pi l}\int_{-\infty}^{\infty} dk
   \frac{A_\pm(k)\,k}{A_{\pm}^{2}(k)\,k^{2}
    +B_{\pm}^{2}(k)}\,\frac{1}{p-k},\label{Aeqn}
\end{equation}
where we have used the asymptotic form of $\widetilde{D}$ 
in Eq.\ (\ref{infraphoton}). 
According to massless photon behavior in the infrared region, 
the integral in Eq.\ (\ref{Aeqn}) shows infrared divergence 
if we naively set $B_{\pm}(p)\equiv0$. 
Besides a spurious photon mass and a gauge fixing parameter 
do not play any role to regularize this infrared divergence 
because of the logarithmic behavior of 
$\widetilde{D}(p)$. We therefore need to leave $B_{\pm}$ 
in the integrand as a cut-off to regularize the infrared divergence 
even for the trivial solutions. 

Let us suppose that $A_{\pm}(p)$ are constant in almost 
all of the momentum regions except for $p\approx0$, 
while $p\,A_{\pm}^{\prime}(p)$ are negligible even 
in $p\approx0$. 
For the large momentum $p$ the integral in Eq.\ (\ref{Aeqn}) 
vanishes and we get $A_{\pm}(\infty)=1$. 
For $p\approx0$ we approximate the fermion propagator as
\begin{equation}
 \frac{1}{A_{\pm}^{2}(k)\,k^{2}+B_{\pm}^{2}(k)}
 \approx \frac{1}{k^{2}+B_{\pm}^{2}(0)},\label{approV}
\end{equation}
which obeys our assumption. Thus Eq.\ (\ref{Aeqn}) reduces to
\begin{equation}
 1-A_{\pm}(0) \approx \frac{\alpha_0}{\pi l}
  \int_{-\infty}^{\infty} dk
   \frac{A_{\pm}(k)}{k^{2}+B_{\pm}^{2}(0)},
\end{equation}
and we obtain \cite{shpagin}
\begin{equation}
 A_{\pm}(0) \approx \left[ 1+\frac{\alpha_0}{|lB_{\pm}(0)|} 
 \right]^{-1}.\label{A(0)}
\end{equation}
Although $A_{\pm}(0)$ reduce to zero for the trivial solutions 
$B_{\pm}(p)\equiv0$, $A_{\pm}(p)\approx1$ are satisfied 
except for the momentum around $p\approx0$ and are consistent 
with the requirement of Ward-Takahashi identity 
under the ladder (bare vertex) approximation. 

Now we solve Eq.\ (\ref{PreGapEq1}) and find the nontrivial solutions 
$B_{\pm}(0)\neq0$ in the symmetric limit $J_{\pm} \to0$. 
Since the integral in Eq.\ (\ref{PreGapEq1}) is dominated at the 
infrared region by the logarithmic behavior of the photon propagator 
in Eq.\ (\ref{infraphoton}), the approximation in Eq.\ (\ref{approV}) 
is also valid and Eq.\ (\ref{PreGapEq1}) is simplified as
\begin{equation}
 B_{\pm}(0) = -\frac{\alpha_0}{\pi l}
  \int_{-\infty}^{\infty} dk
   \frac{B_\pm(k)}{k^{2}+B_{\pm}^{2}(0)}
    \ln|-lk|.\label{Beqn}
\end{equation}
If the above gap equation has nontrivial solutions, they should 
satisfy $|lB_{\pm}(0)|\ll 1$ so as to be consistent with the lowest 
Landau level dominance. 
The smallness of the solutions leads to the result that 
the dominant contribution comes from the infrared region, 
$k\approx0$, in the integral in Eq.\ (\ref{Beqn}). 
Thus we can replace $B_{\pm}(k)$ with $B_{\pm}(0)$ 
and obtain the gap equation:
\begin{equation}
 \omega_\pm = -\frac{\alpha_0}{\pi}
 \int_{-\infty}^{\infty} ds
 \frac{\omega_\pm}{s^2 +\omega_\pm^2}\ln |-s|,
 \label{gapeqn}
\end{equation}
where we introduce dimensionless variables 
$\omega_\pm := l B_\pm (0)$.
This equation has the nontrivial solutions $\pm\omega_s$ given by 
$\omega_s =-\alpha_0 \ln \omega_s$, 
as well as the trivial one \cite{shpagin}. 
Note that the nontrivial solution $\omega_s$ satisfies the condition 
$\omega_s \ll 1$, since $\alpha_0 < 1$ for any $e^2$ and $N$. 
If $\omega_\pm$ appear as one of the nontrivial solutions 
$\pm\omega_s$, $A_{\pm}(0)$ are determined as
\begin{equation}
 A_{\pm}(0)\approx\left[1-\frac{1}{\ln \omega_s}\right]^{-1}\approx1,
\end{equation}
which is consistent with the Ward-Takahashi identity, 
$A_{\pm}(p)\equiv1$, due to the smallness of the nontrivial 
solution $\omega_s$.
\subsection{Classification of the nontrivial solutions}
The obtained gap equations are the same for all three 
theories assigned to $|\kappa| = 0$, $\alpha$, $2\alpha$. 
On the other hand, in order to maintain the self-consistent 
magnetic field for each $\kappa$, the dynamical mass 
$m_d$ and the chemical potential $\mu_d$ have to obey the same 
relation as the explicit breaking parameters $m$ and $\mu$ satisfy 
in Eq.\ (\ref{kappa}) (see Fig.\ \ref{syml}). 
In fact $\omega_{\pm}$ do not necessarily choose 
the nontrivial solutions $\pm \omega_s$. 
Consistent solutions are automatically assigned
to each $\kappa$ such as
\begin{equation}
(\omega_+,\,\omega_-)=\left\{
\begin{array}{ll}
(\omega_s, -\omega_s) & ,\kappa=0 \\
(\omega_s,\,\,0)\,\,{\mbox or}\,\,(0,\,\,-\omega_s)
& ,|\kappa|=\alpha \\
(\omega_s,\,\omega_s) & ,|\kappa|=2\alpha
\end{array}\right.,
\end{equation}
where the replacement of $\omega_s$ with $-\omega_s$ provides another 
set of nontrivial solutions. 
It is intriguing that the solutions for $|\kappa|=\alpha$ appear 
as combinations of a nontrivial solution and a trivial one. 
The dynamical variables, $m_d$ and $\mu_d$, 
which are defined by $m_d:= (\omega_{+}-\omega_{-})/2l$ and 
$\mu_d:=(\omega_{+}+\omega_{-})/2l$, respectively, 
are determined as 
\begin{equation}
(m_d,\,|\mu_d|)=\left\{
\begin{array}{lcl}
(\omega_s/l,\,\,0) & \rm{for} & \kappa=0 \\
(\omega_s/2l,\,\omega_s /2l) & \rm{for} & |\kappa|=\alpha \\
(0,\,\,\omega_s /l) & \rm{for} & |\kappa|=2\alpha
\end{array}\right.. \label{SDsol}
\end{equation}

For $\kappa=0$, fermions acquire their dynamical mass, while 
the vacuum becomes empty ($\mu_d = 0)$. 
The fermion self-energy changes entirely into the dynamical mass 
and reproduces the result in Ref.\ \cite{shpagin}.
For $|\kappa|=2\alpha$, the self-energy is used to occupy 
the lowest Landau level fully with fermions. 
Then there is no mass generation and it supports dynamically 
the result in Ref.\ \cite{hosotani}.
Noteworthy case is $|\kappa|=\alpha$ where the self-energy is shared 
by $m_d$ and $\mu_d$. The vacuum is realized as the lowest Landau 
level half-filled by massive fermions. 
This solution is a new one which spontaneously breaks the U$(2N)$ 
symmetry as well as the Lorentz symmetry at the same time \cite{IK}. 


\section{Spontaneous magnetization and fermion mass generation}

\label{sec:poten}
It still remains unknown whether or not the nontrivial solutions 
to SD gap equation as well as the self-consistent magnetic field 
are energetically more favorable than a trivial solution 
without any magnetic field. 
In order to solve the above problem, we have to investigate 
the effective potential which is obtained as the vacuum 
energy shift owing to the spontaneous breaking of Lorentz symmetry 
and/or U$(2N)$ symmetry. 
Since the solutions, $m_d$ and $\mu_d$, are connected with 
the self-consistent magnetic field $B$, the effective potential 
depends only on $B$ and is given as a function $V(B)$. 
It is composed of four parts including the Maxwell energy;
\begin{equation}
 V(B)=V_{\rm CJT}(B)+V_F (B)+V_P (B)+\frac{B^2}{2}. \label{vtotal}
\end{equation}
$V_{\rm CJT}$ denotes the CJT potential \cite{corn} 
which gives the energy difference between a nontrivial vacuum 
and a trivial one under the presence of the magnetic field. 
$V_F$ $(V_P)$ corresponds to the shift of fermion (photon) 
zero-point energy due to the magnetic field in the symmetric limit. 

In the following we derive the potentials, $V_{\rm CJT}$, $V_F$, 
and $V_P$, separately and investigate stability of vacuum 
in the large $N$ limit for each value of $\kappa$. 
We employ strong coupling expansion with respect to 
$1/\alpha l\sim\sqrt{B/e^3}$ in order to estimate the lowest order 
contribution to $V_P$ based on the expansion of vacuum polarization 
functions in two regimes of momentum, 
which is shown in Section \ref{sec:vacum}.
\subsection{Vacuum energy shift due to fermions: $V_F$}
The vacuum energy in the large $N$ limit is given by the first term 
of effective action in Eq.\ (\ref{eff}). 
It depends on the explicit breaking parameters $(m, \mu)$ 
as well as the magnetic field $B$ such as
\begin{equation}
 \Omega\,E_1 (m,\mu;B):=iN\,{\rm Tr\,Ln}\,S^{-1}, \label{E1:0}
\end{equation}
where $\Omega$ denotes the three-dimensional space-time volume 
and $S^{-1}$ is the inverse of a bare fermion propagator. 
It is nothing but the zero-point energy for fermions coupled with 
the magnetic field $B$. 
The shift of zero-point energy for fermions due to the magnetic field 
is determined in the symmetric limit as
\begin{equation}
 V_F (B):= \lim_{J_\pm\to\,0}\left[ E_1 (m,\mu;B)
  -E_1 (m,\mu;B\to0)\right]. \label{VF0}
\end{equation}

It is convenient to use the condensate ${\cal J}_0$ to derive 
the finite density part of $E_1$ ($\mu\neq0$) \cite{zeitlin2}. 
From the definition (\ref{E1:0}), we notice
\begin{eqnarray}
 \frac{\partial E_1(m,\mu;B)}{\partial \mu} &=& 
  -\Omega^{-1}\,iN\,{\rm Tr}\,[\gamma^0 S] \nonumber \\ &=& 
 -{\cal J}_0 (m, \mu), \label{E1:1}
\end{eqnarray}
where ${\cal J}_0$ is the condensate calculated
in Section \ref{sec:cond} and given by Eq.\ (\ref{cond:42}). 
Integrating both sides over $\mu$, we obtain another form of $E_1$;
\begin{equation}
 E_1 (m,\mu;B)=E_1 (m,0;B)-\int_{0}^{\mu} d\mu^{\prime}
  {\cal J}_0 (m,\mu^{\prime};B). \label{E1:2}
\end{equation}
As to the computation of zero density part, 
the proper time method is used. 
We see from the definition (\ref{E1:0}) that 
\begin{eqnarray}
 \Omega\,E_1 (m,0;B) &=& iN\,{\rm Tr\,Ln}\,[m-i\nabla] \nonumber\\
 &=& \frac{iN}{2}\,{\rm Tr\,Ln}\,[m^2 + \nabla^2], \label{E1:3} 
\end{eqnarray}
where $\nabla:=\gamma^{\mu}D_{\mu}[A^{\rm ext}]$ and we have used 
the relation $\gamma^5 \nabla\gamma^5 =-\nabla$. 
The symbol ${\rm Ln}$ denotes a logarithmic function whose argument 
is given by operators, and ${\rm Tr}$ means a trace over space-time 
coordinates as well as over spinor indices. 
The ${\rm Tr}\,{\rm Ln}$ term is rewritten with 
an integration over the proper time;
\begin{equation}
 {\rm Tr\,Ln}\,[m^2 + \nabla^2]=-\int d^3 x 
  \int_{1/\Lambda^2}^{\infty}\frac{ds}{s}\,{\rm tr}\,
   \langle x|\,e^{-is\,[m^2 + \nabla^2]}|x \rangle. \label{trln}
\end{equation}
Eqs.\ (\ref{fpro0}-\ref{fpro2}) lead to the following relation
\begin{equation}
 \langle x|\,e^{-is\,[m^2 + \nabla^2]}|x \rangle =
  \int\!\!\frac{d^3 k}{(2\pi)^3}\exp\left[
   -is \left(m^2 -k_{0}^2 +\frac{\tan(eBs)}{eBs}{\bf k}^2 
 \right) \right]\left[ 1+\gamma^1 \gamma^2 \tan(eBs) \right]. 
  \label{trln2}
\end{equation}
Applying this relation to Eqs.\ (\ref{E1:3}-\ref{trln}), we obtain, 
after Wick rotation and Gaussian integration, 
\begin{eqnarray}
 E_1 (m,0;B) &=& \frac{N}{4\pi^{3/2}}\,|eB|\int_{1/\Lambda^2}^{\infty}
  \frac{ds}{s^{3/2}}\,e^{-sm^2}\coth(|eB|s) \nonumber\\ &=&
 -\frac{N}{2\pi^{3/2}}\Lambda m^2 -\frac{N}{2\pi}\,|eB|^{3/2}
  \left[ml + 2\sqrt{2}\,\zeta\left( -\frac{1}{2},
   \frac{(ml)^2}{2}+1\right)\right]. \label{E1:4}
\end{eqnarray}
where the function $\zeta (z,q)$ denotes the generalized Riemann zeta 
function which has appeared in Section \ref{sec:cond}. 
The finite density part is calculated from the condensate 
${\cal J}_0$ in Eq.\ (\ref{cond:42}) as
\begin{equation}
 \int_{0}^{\mu} d\mu^{\prime}{\cal J}_0 (m,\mu^{\prime};B) = 
  \frac{N}{2\pi}\,|eB|^{3/2}\left[ l(|\mu|-m)\theta(|\mu|-m)
   +2\sum_{n=1}^{\infty}l(|\mu|-{\cal E}_n)
    \theta(|\mu|-{\cal E}_n)\right]. \label{E1:5}
\end{equation}
Thus the explicit formula for $E_1$ is given by
\begin{eqnarray}
 E_1 (m,\mu;B) &=& -\frac{N}{2\pi^{3/2}}\Lambda m^2 
  -\frac{N}{2\pi}\,|eB|^{3/2}
   \Biggl[ ml + 2\sqrt{2}\,\zeta\left( -\frac{1}{2},
    \frac{(ml)^2}{2}+1\right) \nonumber\\ & & 
 +\,l(|\mu|-m)\theta(|\mu|-m)+2\sum_{n=1}^{\infty}l(|\mu|-{\cal E}_n)
  \theta(|\mu|-{\cal E}_n) \Biggr]. \label{E1:52}
\end{eqnarray}
The fermion vacuum energy in the absence of the magnetic field 
is obtained by taking the weak field limit in Eq.\ (\ref{E1:52}). 
We find
\begin{equation}
  E_1 (m,\mu;B\to0)
  = -\frac{N}{2\pi^{3/2}}\Lambda m^2 +\frac{N}{3\pi}m^3,
  \label{E1:6}
\end{equation}
where we have used the asymptotic formula for 
the zeta function \cite{GR},
\begin{equation}
 \zeta(z,q)\sim\frac{1}{(z-1)q^{z-1}}
 \left[ 1+ \frac{z-1}{2q}+\cdots \right],\quad(q\to\infty).
 \label{asymz}
\end{equation}
We notice that the ultraviolet divergence which appears 
in $B\neq0$ can be completely canceled out by another 
ultraviolet divergence in $B=0$. 

After taking the symmetric limit, $V_F$ is determined as
an ultraviolet finite function of $B$, that is,
\begin{equation}
  V_F (B) = -\frac{N}{4\pi} |eB|^{3/2} 4\sqrt{2}\;\zeta(-1/2),
  \label{VF1}
\end{equation}
which is irrespective of the values of $\kappa$, 
or approaches to the symmetric limit.
\subsection{Vacuum energy shift due to the nontrivial solution: 
$V_{\rm CJT}$}
\begin{figure}
\centerline{\hbox{\psfig{file=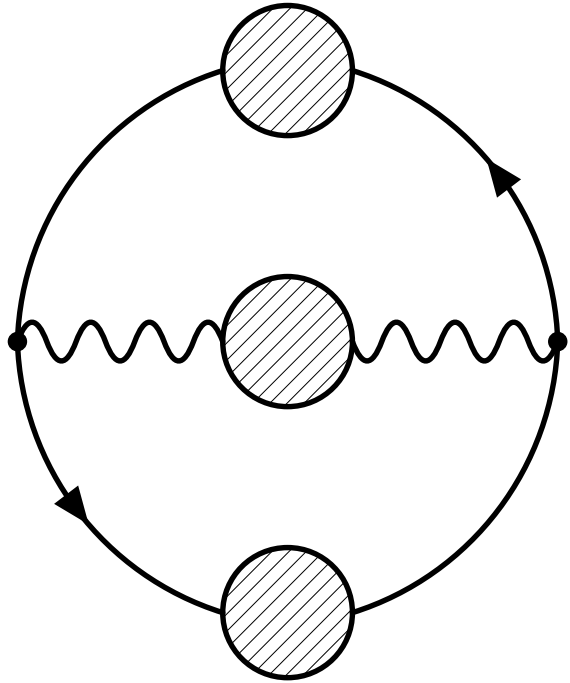,height=9cm}}}\vspace{5mm}
\caption{ The two-particle irreducible (2PI) diagram 
at the leading order in $1/N$ expansion.}\label{2pi}
\end{figure}

The Schwinger-Dyson gap equation is derived from the effective action 
for composite operators, $\bar{\psi}\psi$ and $\psi^\dagger \psi$. 
It is given by a functional $\Gamma[G]$ of the full fermion propagator 
$G$ which is defined by
\begin{equation}
  G(x,y)\,\delta_{ij}
  := i\,\langle 0|T\psi_i (x)\bar{\psi}_j (y)|0 \rangle.
\end{equation}
The equation which leads to stationary points of $\Gamma[G]$ is 
equivalent to the SD gap equation. 
The effective action $\Gamma[G]$, or Cornwall-Jackiw-Tomboulis 
action, is given by \cite{corn}
\begin{equation}
 N\,\Gamma[G]=-iN\,{\rm Tr}\left[ {\rm Ln}\,G^{-1} 
  +S^{-1}G -1 \right] +N\,\Gamma_2 [G], \label{VCJT0}
\end{equation}
where $\Gamma_2 [G]$ denotes all of the 2PI (two-particle irreducible) 
bubble diagrams composed of the full propagator $G$. 
The large $N$ contributions to $\Gamma_2 [G]$ are given by a 
gauge invariant two-loop diagram (see Fig.\ \ref{2pi});
\begin{equation}
 N\,\Gamma_2 [G]=\frac{Ne^2}{2}\,{\rm Tr}
  \left[ G\gamma^\mu G\gamma^\nu \right]D_{\mu\nu}, \label{2PI}
\end{equation}
where the photon propagator $D_{\mu\nu}$ involves the 
large $N$ screening of fermions and has already been 
given in Section \ref{sec:vacum}. 
We see that a stationary point for $\Gamma[G]$ actually provides 
the SD gap equation
\begin{equation}
 \frac{\delta \Gamma[G]}{\delta iG} \equiv
  G^{-1}-S^{-1} +\frac{\delta \Gamma_2 [G]}{\delta iG}=0, \label{stat}
\end{equation}
which has been solved in Section \ref{sec:sd}
and shows the nontrivial solutions 
given by Eq.\ (\ref{SDsol}) as well as a trivial one. 

What we want to know is stability of the vacuum which is realized 
as a nontrivial solution $(m_d, \mu_d)$ for each $\kappa$. 
To see this we only have to investigate which is energetically 
more favorable between the nontrivial solution and the trivial one. 
The vacuum energy due to the dynamical solutions $(m_d, \mu_d)$ is 
provided by the CJT action as its value on the stationary point;
\begin{eqnarray}
 \Omega\,E(m_d,\mu_d;B) &:=&  -N\,\Gamma[G_{\rm sol}] \nonumber\\ 
  &=& iN\,{\rm Tr}\left[ {\rm Ln}\,G_{\rm sol}^{-1} 
   +\frac{1}{2}\left( S^{-1}-G_{\rm sol}^{-1} \right) 
    G_{\rm sol} \right], \label{E0:1}
\end{eqnarray}
where $G_{\rm sol}$ means the solutions to the SD equation 
(\ref{stat}) and we have eliminated the 2PI contribution 
by means of Eq.\ (\ref{stat}). 
Thus the vacuum energy shift due to the nontrivial solution 
is determined as
\begin{equation}
 V_{\rm CJT}(B) := \lim_{J_\pm \to\,0}
  \left[ E(m_d, \mu_d;B) -E(0,0;B) \right]. \label{VCJT1}
\end{equation}
To estimate $V_{\rm CJT}$ from the solutions obtained
in Section \ref{sec:sd}, 
we adopt the following procedure as an approximation, that is 
the replacement
\begin{equation}
  \lim_{J_\pm\to\,0}G_{\rm sol}
  \sim S\,|_{\,(m,\,\mu)\,\to\,(m_d,\,\mu_d)},
  \label{repl}
\end{equation}
where $S$ denotes the bare fermion propagator
defined by Eq.\ (\ref{fpro0}). 
All of the Landau levels contribute to $V_{\rm CJT}$. 
However, the lowest Landau level dominates due to smallness 
of the solutions, $m_d l \ll 1$ and $\mu_d l \ll 1$. 
Under the replacement (\ref{repl}) we see
\begin{equation}
 \lim_{J_\pm\to\,0}iN\,{\rm Tr\,Ln}\,G_{\rm sol}^{-1} = 
  \Omega\,E_1 (m_d,\mu_d;B),
\end{equation}
where the energy function $E_1$ is the same 
as the shift of fermion zero-point energy. 
We also find that the second term in Eq.\ (\ref{E0:1}) becomes
\begin{equation}
 \lim_{J_\pm\to\,0}\frac{iN}{2}\,{\rm Tr\,Ln}
  \left[ (\mu_d \gamma^0 -m_d)\,G_{\rm sol} \right] 
 = \frac{\Omega}{2}\left[ \mu_d \,{\cal J}_0 (m_d,\mu_d) 
  -m_d \,{\cal J}(m_d,\mu_d) \right], 
\end{equation}
where the functions ${\cal J}_0$ and ${\cal J}$ are the same 
condensates calculated in Section \ref{sec:cond}
except for the replacement $(m,\mu)\to(m_d,\mu_d)$. 
Using Eqs.\ (\ref{cond:41},\ref{cond:42}) and Eq.\ (\ref{E1:52}) 
we obtain
\begin{eqnarray}
 \lim_{J_\pm\to\,0}E(m_d,\mu_d;B) &=& -\frac{N}{4\pi}\,|eB|^{3/2}
  \Biggl[ m_d l +l(|\mu_d|-m_d)\,\theta(|\mu_d|-m_d) \nonumber\\
 & & -\sqrt{2}\;(m_d l)^2 
  \,\zeta\left( \frac{1}{2},\frac{(m_d l)^2}{2}+1 \right) 
   +4\sqrt{2}\;\zeta\left( -\frac{1}{2},\frac{(m_d l)^2}{2}+1 \right)
    \Biggr],
\end{eqnarray}
where only the lowest Landau level contributes to the finite density 
part due to smallness of the solutions, while the zero density 
parts appear as a summation over all the Landau levels. 
Notice that the ultraviolet divergences in $E_1$ and ${\cal J}$ 
are completely canceled out in the whole of $E$. 

Thus $V_{\rm CJT}$ is determined in such a form which depends on 
the nontrivial solutions in Eq.\ (\ref{SDsol}) as
\begin{equation}
 V_{\rm CJT}(B) = -\frac{N}{4\pi}\,|eB|^{3/2}
  \biggl[ l \max\{m_d, |\mu_d|\} +O((m_d l)^4) \biggr], \label{VCJT2}
\end{equation}
which shows that the vacuum energy is shifted to be negative. 
Therefore it is confirmed that dynamically generated solutions are 
energetically favorable for each $\kappa$ in the self-consistent 
magnetic field $B$. 
\subsection{Vacuum energy shift due to photon: $V_P$}
$V_P$ is derived from the effective action (\ref{eff}) 
as the next-to-leading-order contribution in $1/N$, that is, 
\begin{eqnarray}
 V_P (B) &:=& \lim_{J_\pm\to\,0}\left(\frac{-1}{2\,\Omega}\right)
  i\,{\rm Tr\,Ln}\,D^{-1}\,-(B\rightarrow 0)\nonumber\\ &=& 
   \lim_{J_\pm\to\,0}\frac{1}{2}\int\!\!\frac{d^3 p}{i(2\pi)^3}
    \,{\rm tr\,Ln}\left[ \widehat{D}(p)D^{-1}(p) \right],
     \label{VP0}
\end{eqnarray}
where we have introduced $\widehat{D}_{\mu\nu}(p)$ 
which is defined as the photon propagator $D_{\mu\nu}(p)$ in $B\to0$, 
and used matrix notation for space-time indices. 
The symbol ${\rm tr}$ thereby means the trace
for space-time indices. 
We notice that $V_P$ is the shift of photon zero-point energy 
due to the magnetic field.$\!\!$
\footnote{
Hereafter we use the propagator $D$ ($\widehat{D}$) where 
the symmetric limit $J_\pm\to 0$ has been taken.}

In order to estimate the integral in Eq.\ (\ref{VP0}), 
we have to expand $V_P$ into a power series with respect 
to the magnetic field $B$. 
However we have supposed the strong magnetic field which 
supports the lowest Landau level dominance and therefore 
guarantees the analysis of the SD gap equation
in Section \ref{sec:sd}. 
In fact we are led to the small solutions, 
$m_d\,(|\mu_d|) \ll l^{-1}$, in comparison with the magnetic field. 
Hence we need another scale which can be taken to be larger than 
the magnetic field and provides the small parameter for the expansion. 

We have indeed such a scale, that is, the gauge coupling $\alpha$ 
to which the Chern-Simons coefficient $\kappa$ is related through 
the Gauss law. The parameter $r:=1/\alpha l$ can be small by 
choosing a sufficiently large $\alpha$ 
with the magnetic length $l$ fixed finite. 
This results in a hierarchy of three scales, 
$m_d\,(|\mu_d|)\ll l^{-1}\ll\alpha$, 
which should be achieved for our present study and the solutions 
$(m_d,\mu_d)$ actually allow this situation. 
Thus we recognize that the expansion of $V_P (B)$ is a kind of 
strong coupling expansion with respect to $1/\alpha$. 
It is enough to pick up and estimate the lowest order contribution of 
the expansion with respect to $r$ in order to investigate 
stability of vacuum. 

To do the above procedure it is necessary at first to expand 
an integrand of Eq.\ (\ref{VP0}) in such a consistent way that 
matches the expansion with respect to $1/\alpha l$. 
In the following we apply an expansion of $D$ around $\widehat{D}$ 
and try to estimate the lowest order contribution to $V_P$. 
The analysis is based on the expansion of vacuum polarization 
in two regimes of momentum in Section \ref{sec:vacum}. 

The propagators $D$ and $\widehat{D}$ are given by
\begin{eqnarray}
 D_{\mu\nu}^{-1}(p) &:=& \Delta_{\mu\nu}^{-1}(p) -\Pi_{\mu\nu}(p), 
  \nonumber\\
 \widehat{D}_{\mu\nu}^{-1}(p) &:=& \Delta_{\mu\nu}^{-1}(p) 
  -\widehat{\Pi}_{\mu\nu}(p), \label{Dpro:1}
\end{eqnarray}
where $\widehat{\Pi}_{\mu\nu}(p)$ denotes the vacuum polarization 
in the limit $B\to0$ and is expressed as$\!\!$
\footnote{
The parity-violating part vanishes in the limit $B\to 0$ 
as shown in Eq.\ (\ref{pola7}) due to our setting of fermions.}
\begin{equation}
 \widehat{\Pi}_{\mu\nu}(p) = (p_\mu p_\nu -p^2 g_{\mu \nu})\,
  \widehat{\Pi}(p), \label{tilpi:1}
\end{equation}
which is written in the gauge invariant form even 
in the presence of the finite density $\mu\neq0$. 
The function $\widehat{\Pi}(p)$ is determined as
\begin{eqnarray}
 \widehat{\Pi}(p) &=& \lim_{J_\pm\to\,0}
  \frac{\alpha}{p}\Bigl[ \theta(m-|\mu|)I(p;m) 
 +\,\theta(|\mu|-m)\,\theta\!\left[ p^2 -4(\mu^2 -m^2) \right]\!
  I(p;\mu) \Bigr] \nonumber\\ &=& 
 \left(\frac{\pi}{2}\right)\frac{\alpha}{p},
\end{eqnarray}
with the function $I(p;u)$ given by Eq.\ (\ref{b:15}). 
From Eq.\ (\ref{Dpro:1}) $D$ is rewritten in the form of 
expansion around $\widehat{D}$ such as
\begin{eqnarray}
 D_{\mu\nu}^{-1}(p) &=& \widehat{D}_{\mu\nu}^{-1}(p)
  -\delta\Pi_{\mu\nu}(p), \nonumber\\
 \delta\Pi_{\mu\nu}(p) &:=& \Pi_{\mu\nu}(p)-\widehat{\Pi}_{\mu\nu}(p).
  \label{delpi}
\end{eqnarray} 
Thus we obtain the following expansion
\begin{eqnarray}
 {\rm tr\,Ln}\left[ \widehat{D}(p)D^{-1}(p) \right] &\equiv& 
 {\rm tr\,Ln}\left[ 1 -\widehat{D}(p)\,\delta\Pi(p) \right]
 \nonumber\\ 
 &=& -\sum_{n=1}^{\infty}\frac{1}{n}\,
 {\rm tr}\left[ \widehat{D}(p)\,\delta\Pi(p) \right]^n.
 \label{exp:1}
\end{eqnarray}
Each term in the above summation is estimated
through the decomposition of a rank 2 $3\times3$ matrix
$\widehat{D}\,\delta\Pi$ into a $2\times2$ unit matrix
and Pauli matrices. It is expressed as
\begin{equation}
 {\rm tr}\left[ \widehat{D}(p)\,\delta\Pi(p) \right]^n 
   = 2\sum_{m=0}^{[n/2]}{}_n C_{2m} \frac{\left[ R(p) \right]^{n-2m}
    \left[ {\bf S}^2 (p) \right]^m}{\left[ Q(p) \right]^n},
\end{equation}
where $[n/2]$ means the integer part of $n/2$ and 
the functions $Q$, $R$, and ${\bf S}=(S_1, S_2, S_3)$
are determined as
\begin{eqnarray}
 Q(p) &=& p^2 \left[ 1 +\widehat{\Pi}(p) \right]^2 +\kappa^2, 
  \nonumber\\
 R(p) &=& -p^2 \left[ 1 +\widehat{\Pi}(p) \right]
  \left[ \delta\Pi_e (p) +\frac{p_{\perp}^2}{2 p^2}\,
   \delta\Pi_\perp (p) \right]
    +\kappa\,\delta\Pi_o (p), \nonumber\\
 S_1 (p) &=& 
  -\left[ 1 +\widehat{\Pi}(p) \right]\,\frac{p_{\perp}^2}{2}\,
   \delta\Pi_\perp (p), \label{RS1}\\
 S_2 (p) &=& \frac{p_{\perp}^2}{2 p}\,
  \kappa\,\delta\Pi_\perp (p), \nonumber\\
 S_3 (p) &=& -ip \left\{ 
  \left[ 1 +\widehat{\Pi}(p) \right]\delta\Pi_o (p)
   +\kappa\left[ \delta\Pi_e (p) +\frac{p_{\perp}^2}{2 p^2}
    \,\delta\Pi_\perp (p) \right] \right\}. \nonumber
\end{eqnarray}
The deviations $\delta\Pi_e$, $\delta\Pi_\perp$, and $\delta\Pi_o$ 
are the counter parts for $\Pi_e$, $\Pi_\perp$, and $\Pi_o$ 
in Eq.\ (\ref{delpi}), respectively. 

In Section \ref{sec:vacum} we have expanded
the vacuum polarization functions in each regime of momentum. 
The energy shift $V_P$ is divided into two parts 
according to two regimes of the loop momentum. 
After the Wick rotation, it is given by
\begin{eqnarray}
 V_P (B) &=& \sum_{n=1}^{\infty}\sum_{m=0}^{[n/2]} 
  \frac{{}_n C_{2m}}{n}
   \left[ V_{nm}^{<}(B) +V_{nm}^{>}(B) \right], \nonumber\\
 V_{nm}^{<}(B) &:=& \frac{-1}{4\pi^2}
  \int_{0}^{\pi} d\theta\sin\theta
   \int_{0}^{l^{-1}} dp\,p^2 \frac{\left[ R(p) \right]^{n-2m}
    \left[ {\bf S}^2 (p) \right]^m}{\left[ Q(p) \right]^n}, 
     \label{VP1}\\
 V_{nm}^{>}(B) &:=& \frac{-1}{4\pi^2}
  \int_{0}^{\pi} d\theta\sin\theta
   \int_{l^{-1}}^{\infty} dp\,p^2 \frac{\left[ R(p) \right]^{n-2m}
    \left[ {\bf S}^2 (p) \right]^m}{\left[ Q(p) \right]^n}, 
     \nonumber
\end{eqnarray}
where $\theta$ denotes the polar angle for 
the Euclidean momentum $p^{\mu}$. 
We apply the (asymptotic) expansion for the vacuum polarization 
functions in the strong (weak) field regime of momentum to 
$V^{<}_{nm}$ ($V^{>}_{nm}$). 

In the strong field regime ($p<l^{-1}$), 
we notice from the results of Eq.\ (\ref{IRc}) that
\begin{equation}
  \delta\Pi_e = \frac{\alpha l}{\sqrt{x}}
  \left[ -\frac{\pi}{2} + \sqrt{x}\,L_e (x) \right],\quad
  \delta\Pi_\perp =\alpha l\, L_{\perp}(x), \quad 
  \delta\Pi_o = \alpha\, L_o (x),
\end{equation}
where $L_e$, $L_{\perp}$, and $L_o$ are expressed
as analytic functions. 
Substituting the above results into Eq.\ (\ref{RS1}), 
$Q$, $R$, and ${\bf S}^2$ 
are written in such a factorized form as
\begin{equation}
  Q = \alpha^2 \left[ \left( \frac{\pi}{2} + r\sqrt{x} \right)^2 
  +\bar{c}^2 \right], \quad
  R = \alpha^2 f\!\left( \sqrt{x}; r \right), \quad 
  {\bf S}^2 = \alpha^4 g\!\left( \sqrt{x}; r \right), \label{RS2}
\end{equation}
where analytic functions $f(u;r)$ and $g(u;r)$ depend on 
the small parameter $r:=1/\alpha l$. 
Then $V_{nm}^{<}$ becomes
\begin{equation}
 V_{nm}^{<}(B) = \frac{-1}{2\pi^2 l^3}
  \int_{0}^{1} d\eta\int_{0}^{1} du\,
   \frac{u^2 \left[ f(u; r) \right]^{n-2m}\left[ g(u; r) \right]^m}
   {\left[ (ru +\frac{\pi}{2})^2 +\bar{c}^2 \right]^n}.
   \label{Vnm:11}
\end{equation}
We notice that a factor $-1/2\pi^2 l^3$ supplies $r^3$. 
If an integral over $u$ diverges by going to $r=0$, it lowers the 
power of $r$ to be less than three. 
Since in the limit $r\to0$ both of $f$ and $g$ are analytic
around $u=0$ and denominators of integrands become constants,
the integrals in Eq.\ (\ref{Vnm:11}) converge and the lowest order 
is determined as $O(r^3)$ which is the same order as 
$V_F$ and $V_{\rm CJT}$. 

In the weak field regime ($p>l^{-1}$) we find from the results of 
Eq.\ (\ref{UVC2}) that
\begin{equation}
  \delta\Pi_e = \frac{\alpha l}{x^{5/2}}\,
  H_e \!\left(\frac{1}{x}\right), \quad
  \delta\Pi_\perp = \frac{\alpha l}{x^{5/2}}\,
  H_{\perp} \!\left(\frac{1}{x}\right),\quad
  \delta\Pi_o = \frac{\alpha}{x}\,
  H_o \!\left(\frac{1}{x}\right),
\end{equation}
where $H_e$, $H_{\perp}$, and $H_o$ are also expressed as 
analytic functions. 
Substituting the above results into Eqs.\ (\ref{RS1}), 
$Q$, $R$, and ${\bf S}^2$ 
are written in such a factorized form as
\begin{equation}
  Q = \alpha^2 x \left[ \left( r +\frac{\pi}{2}\frac{1}{\sqrt{x}} 
  \right)^2 +\frac{\bar{c}^2}{x} \right],\quad
  R = \frac{\alpha^2}{x}\,F\!\left( \frac{1}{\sqrt{x}}; r \right), 
  \quad
  {\bf S}^2 = \frac{\alpha^4}{x}\,
  G\!\left( \frac{1}{\sqrt{x}}; r \right), \label{RS3}
\end{equation}
by using analytic functions $F(u;r)$ and $G(u;r)$. 
Then we obtain the following formula for $V_{nm}^{>}$.
\begin{equation}
 V_{nm}^{>}(B) = \frac{-1}{2\pi^2 l^3}
  \int_{0}^{1} d\eta\int_{0}^{1} du\,
   \frac{u^{4(n-1)}\left[ F(u;r) \right]^{n-2m}
    \left[ G(u;r) \right]^m}{u^{2m}\left[ (r +\frac{\pi}{2}u)^2 
     +\bar{c}^2 u^2 \right]^n}. \label{Vnm:12}
\end{equation}
The convergence at $r=0$ is not so trivial this time because of 
the factor $u^{4(n-1)}/u^{2m}$. 
However $G(u;r)$ supplies another factor $u^2$ when $r=0$, 
which can be seen in the explicit form of it. 
Therefore we obtain an identity $G(u;0)=u^2 \widetilde{G}(u)$ 
with the analytic function $\widetilde{G}(u)$. 
Inserting the above identity into Eq.\ (\ref{Vnm:12}), 
we find in the limit of $r\to0$
\begin{equation}
 V_{nm}^{>}(B) \propto \frac{-1}{ l^3}
  \int_{0}^{1} d\eta\int_{0}^{1} du\,
   u^{2(n-2)}\left[ F(u;0) \right]^{n-2m}[ \widetilde{G}(u) ]^m.
\end{equation}
The lowest order contribution would appear in the above integrals 
if they show divergence. 
The integral shows divergence only when $n=1$, namely, $V_{10}^{>}$. 
In order to extract the lowest order contribution, it is therefore 
sufficient to estimate the coefficient of the most singular part 
of the integral in $V_{10}^{>}$:
\begin{equation}
 V_{10}^{>}(B) = \frac{-1}{2\pi^2 l^3}
  \int_{0}^{1} d\eta\int_{0}^{1} du\,
   \frac{F(u;r)}{(r +\frac{\pi}{2}u)^2 +\bar{c}^2 u^2}.
\end{equation}
From the definition in Eq.\ (\ref{RS3}) the function $F(u;r)$ 
is given by
\begin{equation}
 F(u;r) = \bar{c}\, H_o(u^2) -u\left(r +\frac{\pi}{2}u \right)
  \left[ H_e (u^2) +\frac{\eta^2}{2}H_{\perp}(u^2) \right],
\end{equation}
where we see that the second term supplies a factor $u^2$ at $r=0$ 
so that it never contributes to the lowest order. 
Thus it is clarified that the lowest order contribution is determined 
only by $H_o (0)$ such as
\begin{eqnarray}
 V_{10}^{>}(B) &=& \frac{-1}{2\pi^2 l^3}\,
  \left[ \frac{1}{r}H_o (0)\arctan
   \left( \frac{2\bar{c}}{2r +\pi} \right) +\mbox{const.}
    +O(\ln r) \right], \nonumber\\ &=& 
 -\frac{\alpha}{2\pi^2 l^2}\,H_o (0)
  \arctan \left( \frac{2\bar{c}}{\pi} \right) + O(l^{-3}).
   \label{v10}
\end{eqnarray}
This means that even if we take all of the large $N$ contributions 
in vacuum polarization into account, the lowest order 
contribution to $V_P$ is determined only by $H_o (0)$. 
Therefore the above result is exact in the large $N$ limit. 
The lowest coefficient $H_o (0)$ has already estimated as 
$H_o (0)=2\bar{c}$ which leads the coefficient of $|eB|$ 
to be negative definite.

Thus the vacuum energy shift $V_{P}$ up to $O(B^{3/2})$ is 
obtained as $V_{10}^{>}$ in Eq.\ (\ref{v10}). 
Specifically, it is given by
\begin{equation}
 V_P (B) = -\frac{|\kappa|}{\pi^2}\,|eB|\,\arctan
  \left( \frac{2|\kappa|}{\pi\alpha} \right) +O(B^{3/2}).
\end{equation} 
It is obvious that the negative linear term for $|eB|$ is 
owing to the absence of topological photon mass in $B\neq0$ 
as well as the presence of it in $B=0$ as pointed out by 
Hosotani \cite{hosotani}. 
If we set $\bar{c}=\pm2$, the above result completely matches 
Ref.\ \cite{hosotani}.  
\subsection{Stability of vacuum}
Now we are in a position to discuss the stability of vacuum, 
or a possibility of spontaneous magnetization. 
Recall that the entire effect of vacuum energy shift is given by 
Eq.\ (\ref{vtotal}). 
We notice that the negative linear term for $B$ appears 
only in theories with nonzero $\kappa$, in which the magnetic 
field is possibly supported by the charge condensation 
$e\langle\psi^\dagger \psi\rangle$. 
Therefore, in the theories with $|\kappa|=\alpha$ and $2\alpha$, 
the potential has its stable stationary point at $B\neq0$, 
so the spontaneous magnetization occurs and Lorentz symmetry 
is dynamically broken. 

In the theory with $|\kappa|=2\alpha$, the SD gap equation tells us 
that the fermion self-energy is exhausted to induce a charge density 
on vacuum so that there is no generation of fermion mass. 
Then the vacuum appears as a fully-filled magnetized vacuum
\cite{hosotani}. 
Alternatively, in the theory with $|\kappa|=\alpha$, 
the SD gap equation shows not only the charge condensation 
but also the dynamical fermion mass, whose stability under 
the background magnetic field is supported by the CJT potential. 
Moreover the entire energy shift shows that 
the background field itself is also induced dynamically. 
Thus we are led to the half-filled vacuum in which both of the U$(2N)$ 
symmetry and the Lorentz symmetry are spontaneously broken \cite{IK}.

As to the theory with $\kappa=0$, that is QED${}_3$ 
with $N$ four-component fermions, the situation 
becomes rather subtle because there is no Chern-Simons term, 
or the negative linear term in $B$. 
The lowest order term in $V_P$ is provided as a $B^{3/2}$ term, 
so all of higher-order corrections in the expansion of $\Pi_{\mu\nu}$ 
contribute to its coefficient. 
However, in the entire energy shift $V$,
the $B^{3/2}$ term is saturated by that of $V_F$ with a factor $N$. 
Therefore at least in the large $N$ limit the entire potential has a 
positive $B^{3/2}$ term and results in no spontaneous magnetization. 

We can see this more explicitly in the formula for $V_P$. 
Through an identity ${\rm Tr\,Ln}\equiv\ln{\rm Det}$, 
it is rewritten as
\begin{equation}
 V_P (B) = \lim_{J_\pm\to\,0}\frac{1}{2}\int\!\!\frac{d^3 p}{i(2\pi)^3}
  \,\ln\left[ \frac{[1 +\Pi_e (p)]
   [1 +\Pi_e (p) +(p_{\perp}^2/ p^2)\Pi_{\perp}(p)]}
   {[1 +\widehat{\Pi}(p)]^2} \right],
\end{equation}
in QED${}_3$ [cf. Eq.\ (\ref{det})]. 
Since in QED${}_3$ a photon field is always massless whether a 
magnetic field is turned on or off, there is no essential difference 
due to the magnetic field in the weak field regime of momentum.$\!\!$
\footnote{
This situation changes drastically in Chern-Simons QED${}_3$. 
If $B=0$, a photon has a large topological mass of the order of 
$\alpha$ ($>l^{-1}$) without any screening. 
Therefore the photon behaves as a massive field even in the weak field 
regime of momentum ($p>l^{-1}$) when $B=0$.}
We therefore cut off the integral at $p=l^{-1}$ and approximate 
the vacuum polarization functions as their values at $p=0$. 
Then we obtain
\begin{equation}
 V_P (B) = \frac{1}{4\pi^2 l^3}\int_{0}^{1} d\eta\int_{0}^{1}
  du\,u^2 \ln \left[ \frac{[r +c][r +c -c\,\eta^2 /2]}
   {[r +\pi/2u]^2} \right].
\end{equation}
We see that the integrand becomes negative
around $u=0$ due to the $1/p$ 
behavior of $\widehat{\Pi}(p)$ so that $V_P$ is led to be negative. 
The integrals can be performed analytically
and $V_P$ is obtained as 
\begin{equation}
V_P (B) = \frac{1}{12\pi^2}|eB|^{3/2} 
 K\!\left( \frac{2}{\pi\alpha}\sqrt{|eB|} \right),
\end{equation}
with the function $K(x)$ defined by
\begin{eqnarray}
 K(x) &:=& -\frac{8}{3}-\frac{2}{x^3}
  \left[ \ln(x+1)-x+\frac{x^2}{2}-\frac{x^3}{3} \right] 
 +\ln \left( \frac{x+b}{x+1} \right) 
  +\ln \left( \frac{x+b/2}{x+1} \right) \nonumber\\
 & & +\; 2\sqrt{\frac{2(x+b)}{b}}{\rm arccoth}\sqrt{\frac{2(x+b)}{b}},
\end{eqnarray}
where we have introduced $b:=2c/\pi\simeq 1.12298$. 

\begin{figure}
\centerline{\hbox{\psfig{file=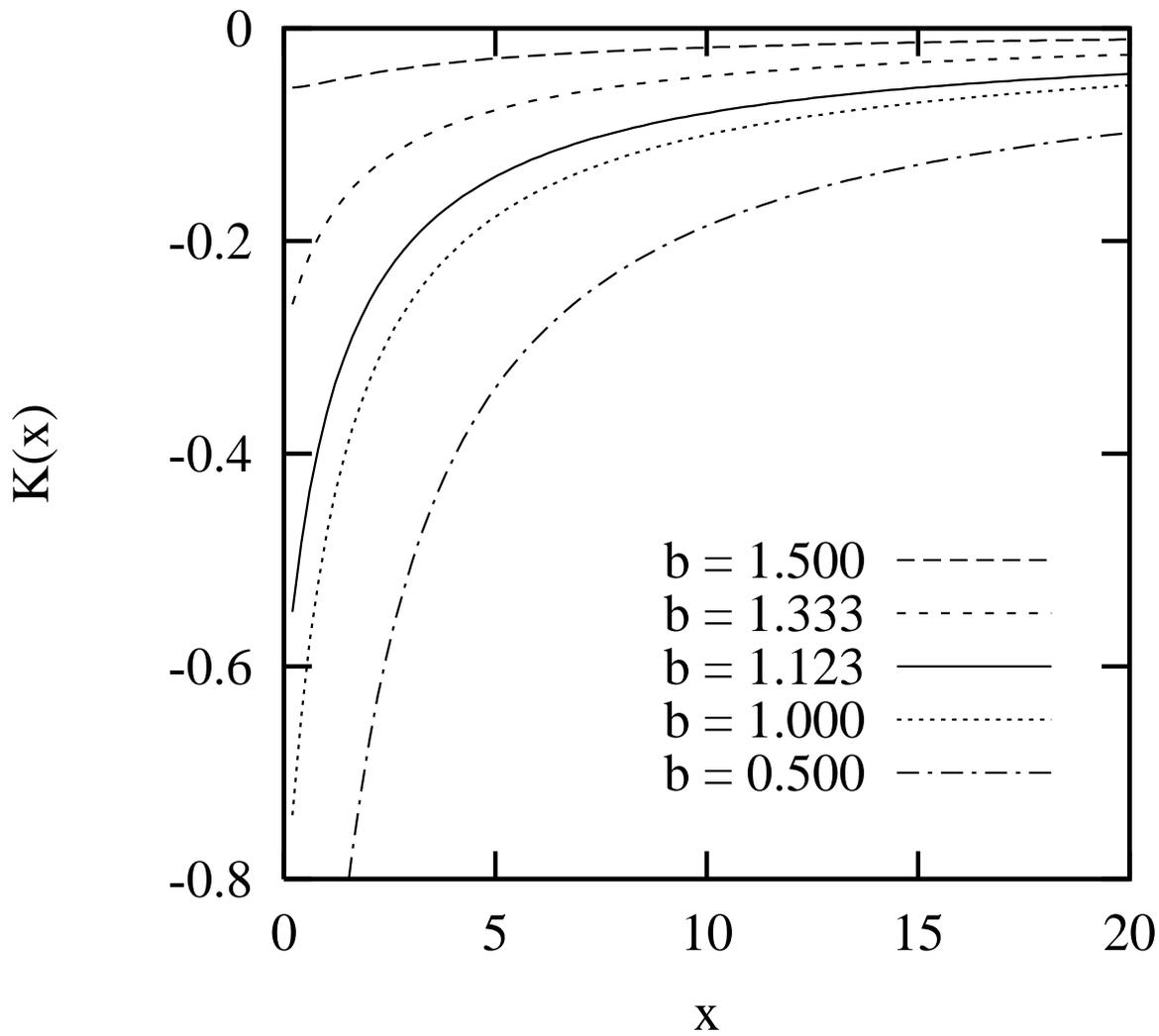,height=15cm}}}\vspace{5mm}
\caption{ The $b$ dependence of the function $K(x)$.}\label{k-b}
\end{figure}

$K(x)$ shows monotonically increasing behavior and becomes 
negative definite in $x\ge0$ for $b<4/3$, that is $K(0)\le K(x)<0$ 
as shown in Fig.\ \ref{k-b}. 
Hence, if an inequality $|K(0)|< 2\pi cN$ is satisfied, then the 
$B^{3/2}$ term in the entire energy shift $V(B)$ becomes positive 
semi-definite and shows monotonically increasing behavior which leads 
us to a non-magnetized vacuum. 
If we suppose large $N$, the above inequality is trivially 
satisfied, which only means that the $B^{3/2}$ term is dominated 
by the large $N$ contribution of $V_F$. 
However the parameter $c$ (or $b$), which is determined 
by the kinematics of fermion in the magnetic field, satisfies
\begin{eqnarray}
 \frac{|K(0)|}{2\pi c} &=& \frac{1}{2\pi c}\left| -\frac{8}{3}
  +2\sqrt{2}\,{\rm arccoth}\sqrt{2}
   +\ln(2c^2 /\pi^2) \right| \nonumber\\ 
 &\simeq& 0.0572877.
\end{eqnarray}
Therefore, even if $N=1$, the spontaneous magnetization
or equivalently the spontaneous Lorentz symmetry breaking 
does not occur in QED${}_3$ with $N$ four-component fermions. 

\section{Conclusion}

\label{sec:concl}
In this paper we have investigated dynamical symmetry breaking 
in Chern-Simons QED${}_3$ associated with the realization of 
finite density vacua (or occupied lowest Landau levels). 

Through the explicit estimation of condensates 
$\langle\bar{\psi}\psi\rangle$ and $\langle\psi^\dagger \psi\rangle$, 
we have clarified that in the presence of the Chern-Simons 
interaction the magnetic field is necessarily connected to 
the charge condensation $e\langle\psi^\dagger \psi\rangle$ 
through the Gauss law in such a way as to restrict the possible values 
of the CS coefficient $\kappa$ to $\pm Ne^2 /2\pi$, 
$\pm Ne^2 /4\pi$, and $0$ \cite{hosotani,hosotani2,NS}. 
In other words, the magnetic field can penetrate the system only if 
the photon effectively becomes massless \cite{hosotani}. 
We also have found that both condensates
$\langle\bar{\psi}\psi\rangle$ and
$\langle\psi^\dagger \psi\rangle$ always appear so as to complement 
each other and to keep the combination $\langle\bar{\psi}\psi\rangle
-|\langle\psi^\dagger \psi\rangle|$ constant. 
The symmetry breaking patterns, or the vacuum configurations, 
are displayed in Table.\ I. 

In the theory with $|\kappa|=Ne^2 /2\pi$, the vacuum is realized as 
a fully-filled lowest Landau level [$(\nu_+,\nu_-)=(1,0)$, or $(0,1)$] 
which is a singlet of the flavor U$(2N)$ group. 
Only the Lorentz symmetry is spontaneously broken 
by the induced magnetic field together with charge condensation, 
as pointed out in Refs.\ \cite{hosotani,IS}. 
Alternatively, in the theory with $\kappa=0$, or QED${}_3$ with $N$ 
four-component fermions, the vacuum is realized as an empty lowest 
Landau level [$\nu_\pm =0$] even in the presence of
the magnetic field. 
The U$(2N)$ symmetry is spontaneously broken, while the magnetic 
field is not accompanied by charge condensation \cite{gusynin,shpagin}. 
As to the theory with $|\kappa|=Ne^2 /4\pi$, both condensates coexist 
in the half-filled lowest Landau level 
[$(\nu_+,\nu_-)=(1/2,0)$, or $(0,1/2)$]. 
This vacuum allows the spontaneous breaking of both symmetries, 
that is, the U$(2N)$ symmetry and the Lorentz symmetry \cite{IK}. 

The above situation suggests that in the fermion self-energy 
the scalar component $m_d$ and the $\gamma^0$ component $\mu_d$ 
also complement each other. 
We have also attempted to find the dynamical solutions which 
correspond to the possible configurations of the above three classes 
through the analysis of the Schwinger-Dyson gap equation. 

In the theory with $|\kappa|=Ne^2 /2\pi$, the realization of 
a fully-filled vacuum is supported by the solution 
$(m_d,|\mu_d|)=(0,\omega_s /l)$. 
The fermion self-energy has only a $\gamma^0$ component, 
the originally massless fermions remain massless and 
there is no generation of fermion mass. 
This solution reproduces and verifies the result of
Ref.\ \cite{hosotani}, that is, 
vanishing fermion mass and broken Lorentz symmetry. 
Thus, the magnetic field does not necessarily 
lead to mass generation if the vacuum is fully-filled by fermions.
This is in sharp contrast to the results of Ref.\ \cite{shpagin}, 
which are based on the empty vacuum 
and correspond to the theory with $\kappa=0$. 
In this case the solution to the gap equation has 
$(m_d,|\mu_d|)=(\omega_s /l,0)$. 
The fermion self-energy is saturated by the dynamically 
generated fermion mass, which then results in an empty vacuum. 

In the case $|\kappa|=Ne^2 /4\pi$, we have found a novel
solution $(m_d,|\mu_d|)=(\omega_s /2l,\omega_s /2l)$ which causes 
spontaneous magnetization as well as the dynamical generation of
fermion mass \cite{IK}. 
The scalar component $m_d$ and the $\gamma^0$ component
$\mu_d$ are comparable to each other. 
The fermion self-energy is shared, half filling 
the lowest Landau level and half supplying fermions
their dynamical mass. 
In the resultant half-filled vacuum the U$(2N)$ symmetry 
and the Lorentz symmetry are simultaneously broken 
through the dynamically generated fermion mass and 
the induced magnetic field \cite{IK}. 

We have investigated the question of vacuum stability 
for each class through the explicit calculation of the energy 
shift $V(B)$ due to the magnetic field $B$. 
We find that a negative linear term in $B$ always appears 
for $|\kappa|=Ne^2 /2\pi$ and $Ne^2/4\pi$, leading to a nonzero 
magnetic field at the stationary point. 
Moreover, the estimated coefficient for the above term is 
exact in the large $N$ limit and matches the former 
result in Ref.\ \cite{hosotani}. 

For $\kappa=0$, or QED${}_3$ with $N$ four-component fermions, 
the lowest term in the energy shift behaves as $B^{3/2}$. 
The fermions and the photon make opposite contributions. 
Since the energy shift due to the fermions is the leading one 
at large $N$ and is positive, the entire energy shift becomes 
positive at least as $N\to\infty$. 
However, our rough estimate has shown that even if $N=1$ 
the entire shift is positive due to the kinematics of fermions 
in the magnetic field. 
Therefore we can conclude that spontaneous magnetization does not 
occur in QED${}_3$ with $N$ four-component fermions.$\!\!$
\footnote{Some authors \cite{zeitlin2,AH} claim that spontaneous 
magnetization occurs in QED${}_3$. Their analysis is, however, 
based on two-component fermions and assumes a heavy fermion mass 
which breaks parity explicitly.}

It is an open problem whether or not the magnetized vacua 
found at zero temperature are maintained also at finite 
temperature, although the Lorentz symmetry is explicitly broken 
by the heat bath. The stability of the magnetized vacua at finite 
temperature was confirmed in Ref.\ \cite{kanemura} 
in the same context as Ref.\ \cite{hosotani}. 
On the other hand, Ref.\ \cite{das} found that in $2+1$ dimensions 
and in the external magnetic field the condensate 
$\langle\bar{\psi}\psi\rangle$ is unstable 
at finite temperature $T$ due to nonanalyticity at the origin 
in the $(m,T)$ plane. 
Thus one needs to take into account the possibility of dynamical 
generation of fermion mass as well as spontaneous magnetization 
also in the finite temperature. 
Another problem is to extend our study to non-Abelian gauge theories, 
namely, Chern-Simons QCD${}_3$, whose intimate relationship 
with the frustrated Heisenberg antiferromagnets is studied 
in Ref.\ \cite{ichinose}. 

The gauge sector of Chern-Simons QED${}_3$ is dual to 
the Abelian Chern-Simons-Higgs theory, where the kinetic term 
for the Higgs field provides the gauge boson mass term 
through unitary gauge fixing \cite{DY}. 
On the other hand, the $(2+1)$-dimensional Thirring model can be 
reformulated as an Abelian Higgs theory with the realization of 
a massive composite gauge boson through the introduction 
of a spurious Higgs boson \cite{IKSY}. 
It may, therefore, also be interesting to study 
another infrared sensitive theory;
\begin{equation}
{\cal L}^{\prime}=\bar{\psi}\gamma^{\mu}
  \left[ i\partial_\mu +e{\cal A}_\mu \right]\psi
   -\frac{1}{2G}{\cal A}_\mu {\cal A}^\mu
    -\frac{\kappa}{2}\epsilon_{\mu \nu \rho}{\cal A}^{\mu}
      \partial^{\nu}{\cal A}^{\rho},
\end{equation}
whose dynamical gauge boson might play an important role in planar 
condensed matter systems.

\section*{Acknowledgment}

We would like to thank Koichi Yamawaki and A. I. Sanda 
for their encouragements and enlightening discussions. 
We are grateful to A. Chodos and Yoonbai Kim for his critical 
reading of our manuscript and helpful advice. 
We thank V. P. Gusynin, Michio Hashimoto, 
Kei-Ichi Kondo, V. A. Miransky, Yukinori Nagatani, Toshiro Sato, 
Akihiro Shibata, Masaki Sugiura, and Akira Takamura 
for valuable discussions. 
T. I. is grateful to Deog Ki Hong, Kenzo Inoue, Taro Kashiwa, and 
Koji Harada for their encouragements and useful discussions. 
He is also grateful to Sung Soo Kim, Han-Oh Lee, Tae-Oh Park, 
Dong Hyun Park, and Woo Jin Yang for interesting discussions. 
This work was supported in part by KOSEF Postdoctoral Fellowship 
and KRF (1998-015-D00075).

\appendix
\section{Contour integrals}
\label{sec:appenA}
In this appendix we provide formulas for contour integrals 
which appear in Section \ref{sec:cond} and
Section \ref{sec:vacum}. They are given for 
an arbitrary positive constant $M$ and an integer $n$ by
\begin{eqnarray}
 I_{+}^{(n)}(M) &:=& \frac{1}{\pi}\oint_{C_\mu} dz\,
  \frac{1}{(z^2 + M^2)^n}, \label{cont:11}\\
 I_{-}^{(n)}(M) &:=& \frac{1}{\pi}\oint_{C_\mu} dz\,
  \frac{iz}{(z^2 + M^2)^n}, \label{cont:12}\\
 I^{(0)}(M) &:=& \frac{1}{\pi}\oint_{C_\mu} dz\,
  \ln l^2 (z^2 +M^2), \label{cont:13}
\end{eqnarray}
where the contour $C_\mu$ is a rectangle which connects two 
trajectories $(-R-i\mu,+R-i\mu)$ and $(-R,+R)$, 
and finally we take the limit $R\to\infty$ (see Fig.\ \ref{cont}). 

When $|\mu|=M$, poles or branch points of the integrand 
are put just on the contour $C_\mu$. We therefore replace
the integrals in $|\mu|=M$ as their Cauchy's principal values. 
Notice that integrals with $n\ge2$ show divergence 
at $|\mu|=M$ and become ill-defined. 
To avoid these ambiguities we define them
as $n$-th $M$ derivatives of the integrals $I_{\pm}^{(1)}$, namely,
\begin{equation}
 I_{\pm}^{(n)}(M) = \frac{1}{(n-1)!}
  \left(\frac{-1}{2M}\frac{\partial}{\partial M}\right)^{n-1} 
   I_{\pm}^{(1)}(M). \label{cont2}
\end{equation}
The integrals $I_{\pm}^{(1)}$ are calculated by means of the 
residue theorem. When $|\mu|>M$, the poles $z=\pm iM$ 
of the integrand are located inside the contour $C_\mu$. 
Thus we obtain the following results
\begin{eqnarray}
 I_{+}^{(1)}(M) &=& -\frac{1}{M}\,\theta(|\mu|-M), 
  \label{cont:31}\\
 I_{-}^{(1)}(M) &=& -{\rm sgn}(\mu)\,\theta(|\mu|-M), 
  \label{cont:32} 
\end{eqnarray}
where the step functions take a value $1/2$ due to the redefinition 
as their Cauchy's principal value at $|\mu|=M$ and they match 
the zero temperature limit of Fermi-Dirac distribution function. 

By use of Eq.\ (\ref{cont2}) the integrals $I_{\pm}^{(2)}$ are 
determined as
\begin{eqnarray}
 I_{+}^{(2)}(M) &=& -\frac{1}{2M^3}\,\theta(|\mu|-M)
  -\frac{1}{2M^2}\,\delta(|\mu|-M), \label{cont:41}\\
 I_{-}^{(2)}(M) &=& -\frac{1}{2M}\,{\rm sgn}(\mu)\,
  \delta(|\mu|-M), \label{cont:42} 
\end{eqnarray}
which show delta function singularities at $|\mu|=M$ 
as mentioned above. They cannot be neglected because they appear 
as the zero temperature limit of the $M$ derivative of 
Fermi-Dirac distribution function. 

As to the integral $I^{(0)}$, its integrand does not have 
any pole but branch points at $z=\pm iM$. 
When $|\mu|>M$, the branch points are located inside the contour $C_\mu$ 
therefore the integral is replaced with the one integrated 
along a minimal contour that rounds the branch cut
$|{\rm Im} z|\geq M$ 
on the imaginary axis. Thus we obtain
\begin{equation}
 I^{(0)}(M) = 2(|\mu|-M)\,\theta(|\mu|-M). \label{cont:5}
\end{equation}

\section{Asymptotic expansion}
\label{sec:appenB}
In the following we provide the asymptotic expansion 
for the function $I(X)$ which is given by the integral
\begin{equation}
  I(X):= \int_{0}^{\infty} d\tau\;G(\tau)\,e^{-X\Phi(\tau)}.
  \label{b1}
\end{equation}
We suppose that the function $\Phi(\tau)$ is analytic at $\tau=0$, 
as well as the function $G(\tau)$, and an odd function for $\tau$ 
which monotonically increases in $\tau\ge0$. 
For convenience, we normalize $\Phi$ as
\begin{equation}
 \left.\frac{d}{d\tau}\Phi(\tau)\right|_{\tau=0}=1. \label{b1.5}
\end{equation}
Our aim is to expand $I(X)$ into the asymptotic series 
for the large $X$, that is
\begin{equation}
 I(X)\sim\sum_{n=0}^{\infty}\;\frac{I_n}{X^n},\quad(X\to\infty). 
  \label{b2}
\end{equation}

We notice that there exists the inverse function $f(y):=\Phi^{-1}(y)$ 
which is also an odd function of $y$ and monotonically increases 
in $y\ge0$ from the definition of $\Phi(\tau)$. 
We exchange the variable of integration with $y=\Phi(\tau)$ 
in Eq.\ (\ref{b1}) and obtain
\begin{equation}
 I(X)=\int_{0}^{\infty} dy\,\frac{df(y)}{dy}\,G(f(y))\,e^{-Xy}.
  \label{b3}
\end{equation}
Because of an exponential factor $e^{-Xy}$, 
the neighborhood of $y=0$ dominantly contributes to the integral 
for the large $X$. 
It is therefore allowed in Eq.\ (\ref{b3}) to expand 
\begin{equation}
 F(y):=\frac{df(y)}{dy}\,G(f(y)) \label{b4}
\end{equation}
into a power series of $y$. Thus the asymptotic series of $I(X)$ 
is obtained as
\begin{equation}
 I(X)\sim\frac{1}{X}\sum_{n=0}^{\infty}\,\frac{F^{(n)}}{X^n},
  \quad(X\to\infty),\label{b5}
\end{equation}
where $F^{(n)}$ denotes the $n$-th derivative of $F(y)$ at $y=0$. 
The coefficients $F^{(n)}$ are determined by derivatives of 
$G$ and $f$, and the series up to $O(X^{-4})$ becomes
\begin{equation}
 I(X) \sim \frac{G^{(0)}}{X} + \frac{G^{(1)}}{X^2}
  +\frac{G^{(2)} + f^{(3)} G^{(0)}}{X^3}
   +\,\frac{G^{(3)} + 3f^{(3)} G^{(1)}}{X^4} +\cdots,
    \quad(X\to\infty). \label{b7}
\end{equation}
where $G^{(n)}$ ($f^{(n)}$) denotes the $n$-th derivative of $G$ ($f$) 
at $\tau=0$ ($y=0$). 

Now we apply the above expansion to the vacuum polarization functions 
given in Section \ref{sec:vacum}. 
The functions $\Pi_e$, $\Pi_\perp$, and $\Pi_o$ are written in the 
same form as Eq.\ (\ref{b1}), that is,
\begin{eqnarray}
 \Pi_e (p) &=& \frac{\alpha l^2}{\pi}\int_{0}^{1} dv
  \int_{-\infty}^{\infty} d\bar{k}\int_{0}^{\infty} d\tau\;
   G_e (\tau)\, e^{-X(\bar{k}-i\mu)\Phi(\tau)}, \nonumber\\
 \Pi_\perp (p) &=& \frac{\alpha l^2}{\pi}\int_{0}^{1} dv
  \int_{-\infty}^{\infty} d\bar{k}\int_{0}^{\infty} d\tau\;
   G_\perp (\tau)\, e^{-X(\bar{k}-i\mu)\Phi(\tau)}, \label{b8}\\
 \Pi_o (p) &=& \frac{\alpha l^2}{\pi}\int_{0}^{1} dv
  \int_{-\infty}^{\infty} d\bar{k}\;
   i(\bar{k}-i\mu)\!\int_{0}^{\infty} d\tau\;
    G_o (\tau)\, e^{-X(\bar{k}-i\mu)\Phi(\tau)}, \nonumber
\end{eqnarray}
where we define the functions $X(\bar{k})$ and $\Phi(\tau)$ by
\begin{eqnarray}
 X(\bar{k}) &:=& l^2 \left[ \bar{k}^2 +M_{v}^2 \right],\quad
  M_v := \sqrt{m^2 +\frac{1-v^2}{4}p^2}, \nonumber\\
 \Phi(\tau) &:=& \tau \left\{ 1 + \frac{x}{X(\bar{k}-i\mu)}
  \left[ \frac{\cosh \tau -\cosh \tau v}{2\tau\sinh \tau}
   -\frac{1-v^2}{4} \right]\cos^2 \theta \right\},\label{b10}
\end{eqnarray}
where $x:= (lp)^2$ and $\theta$ denotes the polar angle for the 
Euclidean momentum $p^{\mu}$. 
The functions $G_e$, $G_\perp$, and $G_o$ are read from 
Eq.\ (\ref{pola7}) as
\begin{eqnarray}
 G_e (\tau) &=& \frac{\tau(\cosh \tau v - v\coth \tau\sinh \tau v)}
  {\sinh \tau},\nonumber\\ 
 G_\perp (\tau) &=& \frac{2\tau(\cosh \tau -\cosh \tau v)}
  {\sinh^3 \tau} - G_e (p),\label{b11}\\
 G_o (\tau) &=& \frac{2\tau(\cosh \tau \cosh \tau v -1)}{\sinh^2 \tau}
  \;{\rm sgn}(eB),\nonumber
\end{eqnarray}
all of which are analytic at $\tau=0$.

We notice that $\Phi(\tau)$ is a monotonically increasing odd 
function of $\tau$ and satisfies the normalization condition of
Eq.\ (\ref{b1.5}). 
Therefore the vacuum polarization functions are expanded 
into the same form as the asymptotic series in Eq.\ (\ref{b7}) 
in the weak field regime ($l^{-2}\ll p^2$);
\begin{eqnarray}
 \Pi_e (p) &=& \frac{\alpha l^2}{\pi}\int_{0}^{1} dv
  \int_{-\infty}^{\infty} d\bar{k}\left[ 
   \frac{G_{e}^{(0)}}{X(\bar{k}-i\mu)}
    +\frac{G_{e}^{(1)}}{X^2 (\bar{k}-i\mu)}
     +O(X^{-3}) \right], \nonumber\\
 \Pi_\perp (p) &=& \frac{\alpha l^2}{\pi}\int_{0}^{1} dv
  \int_{-\infty}^{\infty} d\bar{k}\left[ 
   \frac{G_{\perp}^{(0)}}{X(\bar{k}-i\mu)}
    +\frac{G_{\perp}^{(1)}}{X^2 (\bar{k}-i\mu)}
     +O(X^{-3}) \right], \label{b12}\\
 \Pi_o (p) &=& \frac{\alpha l^2}{\pi}\,{\rm sgn}(eB)
  \int_{0}^{1} dv\int_{-\infty}^{\infty}
   d\bar{k}\;i(\bar{k}-i\mu)\left[ 
   \frac{G_{o}^{(0)}}{X(\bar{k}-i\mu)}
    +\frac{G_{o}^{(1)}}{X^2 (\bar{k}-i\mu)}
     +O(X^{-3}) \right]. \nonumber
\end{eqnarray}
Since the only dimensionful parameter is the momentum $p$ 
after taking the symmetric limit $J_\pm\to0$, the $X^{-n}$ terms 
in the above equations are identified with the $n$-th terms of 
the asymptotic series in Eq.\ (\ref{pola13}), 
whose coefficients up to $O(X^{-1})$ are given by
\begin{eqnarray}
 H_{e}^{(0)} &=& 
  \lim_{J_\pm\to\,0} p \int_{0}^{1} dv\,(1-v^2)\,
   \frac{1}{\pi}\left[ \int_{-\infty}^{\infty} dz\,
    \frac{1}{z^2 +M_{v}^2} + \oint_{C_\mu} dz\,
     \frac{1}{z^2 +M_{v}^2} \right], \nonumber\\
 H_{o}^{(1)} &=& \lim_{J_\pm\to\,0}
  p^2 \,{\rm sgn}(eB)\int_{0}^{1} dv\,(1+v^2)\,
   \frac{1}{\pi}\oint_{C_\mu} dz\,
    \frac{iz}{\left[z^2 +M_{v}^2\right]^2}, \label{b13}\\
 H_{\perp}^{(0)} &\equiv& 0. \nonumber
\end{eqnarray}
The other coefficients up to $O(X^{-1})$ become trivially 
zero due to even or odd property of the functions
$G_e$, $G_\perp$, and $G_o$ under the parity. 
The integrals over $z$ are carried out as the contour integrals 
given in Appendix \ref{sec:appenA} and are given by
\begin{eqnarray}
 H_{e}^{(0)} &=& \lim_{J_\pm\to\,0} p
   \int_{0}^{1} dv\,\frac{1-v^2}{M_{v}}
    \left[1-\theta(|\mu|-M_{v})\right], \nonumber\\
 H_{o}^{(1)} &=& -\lim_{J_\pm\to\,0}
  p^2 \,{\rm sgn}(\mu eB)\int_{0}^{1} dv\,\frac{1+v^2}{2 M_{v}}\,
    \delta(|\mu|-M_{v}), \label{b14}
\end{eqnarray}
which have appeared in Section \ref{sec:vacum} and the integrals 
over parameter $v$ can be calculated by using formulas
in Appendix \ref{sec:appenC}.  
Thus we obtain
\begin{eqnarray}
 H_{e}^{(0)} &=& \lim_{J_\pm\to\,0} \Bigl[ \theta(m-|\mu|)I(p;m) 
  +\,\theta(|\mu|-m)\,\theta\!\left[ p^2 -4(\mu^2 -m^2) \right]\!
   I(p;\mu) \Bigr], \nonumber\\
 H_{o}^{(0)} &=& -\lim_{J_\pm\to\,0}{\rm sgn}(\mu eB)\,
  \theta(|\mu|-m)\,\theta\!\left[ p^2 -4(\mu^2 -m^2) \right]
   \frac{4\left[ p^2 -2(\mu^2 -m^2) \right]}
    {p\sqrt{p^2 -4(\mu^2 -m^2)}},
\end{eqnarray}
with the function $I(p;u)$ which is defined as
\begin{equation}
 I(p;u):= \frac{2|u|\sqrt{p^2 -4(u^2 -m^2)}}{p^2}
  +\left( 1 -\frac{4m^2}{p^2} \right) \arctan
   \left( \frac{\sqrt{p^2 -4(u^2 -m^2)}}{2|u|} \right). \label{b:15}
\end{equation}

\section{Feynman integrals at finite densities}
\label{sec:appenC}
In this paper we have used the following integral formulas 
in calculations of loop diagrams;
\begin{eqnarray}
  J^{0}_n (p)
  &:=& \frac{p}{2}\int_{0}^{1} \frac{dv}{M_{v}^n}=I_n (m),\\ 
 J_n (p) &:=& \frac{p}{2}\int_{0}^{1} \frac{dv}{M_{v}^n}\,
  \theta(|\mu|-M_v) \nonumber\\ &=& 
   \theta(|\mu|-m) \Bigl\{ I_n (m) 
    -\theta\!\left[ p^2 -4(\mu^2 -m^2) \right]\!I_n (\mu) \Bigr\},\\
 \bar{J}_n (p) &:=& \frac{p}{2}\int_{0}^{1} \frac{dv}{M_{v}^n}\,
  \delta(|\mu|-M_v) \nonumber\\ &=& 
   \theta(|\mu|-m)\,\theta\!\left[ p^2 -4(\mu^2 -m^2) \right]
    \bar{I}_n(\mu),
\end{eqnarray}
where $n$ is an arbitrary integer and the functions $I_n$ and
$\bar{I}_n$ are given by
\begin{eqnarray}
 I_n (u) &:=& \frac{\sqrt{C(p)-u^2}}{[C(p)]^{n/2}}\,
  {}_2 F_1 \!\left( \frac{1}{2},\frac{n}{2},\frac{3}{2};
   1-\frac{u^2}{C(p)} \right), \\
 \bar{I}_n (u) &:=& \frac{1}{|u|^{n-1}\sqrt{C(p)-u^2}},
\end{eqnarray}
with the hypergeometric function ${}_2 F_1 (a,b,c;z)$ and
\begin{equation}
 C(p) := m^2 +\frac{p^2}{4}.
\end{equation}
For instance the vacuum polarization function $\widehat{\Pi}(p)$ 
in Eq.\ (\ref{tilpi:1}) is given by
\begin{eqnarray}
 \widehat{\Pi}(p) &=& \alpha \int_{0}^{1} dv\,
  \frac{1-v^2}{M_{v}}\left[1-\theta(|\mu|-M_{v})\right] \nonumber\\
 &=&\frac{8\alpha}{p^3}\left\{ J^{0}_{-1}(p)-m^2 J^{0}_1 (p)
  -\left[ J_{-1}(p)-m^2 J_1 (p) \right] \right\} \nonumber\\
 &=& \frac{8\alpha}{p^3}\Bigl\{ \theta(m-|\mu|)
  \left[  I_{-1}(m)-m^2 I_1 (m) \right] \nonumber\\ & & 
   +\,\theta(|\mu|-m)\theta\!\left[ p^2 -4(\mu^2 -m^2) \right]
    \left[ I_{-1}(\mu)-m^2 I_1 (\mu) \right] \Bigr\}.
\end{eqnarray}
The functions $I_{-1}$ and $I_1$ are calculated as
\begin{eqnarray}
 I_{-1}(u) &=& \frac{|u|\sqrt{C(p)-u^2}}{2} +\frac{C(p)}{2}
  \arctan \left[ \frac{\sqrt{C(p)-u^2}}{|u|} \right], \\
 I_1 (u) &=& \arctan \left[ \frac{\sqrt{C(p)-u^2}}{|u|} \right].
\end{eqnarray}
Finally we obtain
\begin{equation}
 \widehat{\Pi}(p) = \frac{\alpha}{p} \Bigl\{ \theta(m-|\mu|)I(p;m) 
  +\,\theta(|\mu|-m)\,\theta\!\left[ p^2 -4(\mu^2 -m^2) \right]\!
   I(p;\mu) \Bigr\},
\end{equation}
where we define the function $I(p;u)$ as
\begin{eqnarray}
 I(p;u) &:=& \frac{8}{p^2}\left[ I_{-1}(u)-m^2 I_1 (u) \right] 
  \nonumber\\ &=& \frac{2|u|\sqrt{p^2 -4(u^2 -m^2)}}{p^2}
   +\left( 1 -\frac{4m^2}{p^2} \right) \arctan
    \left( \frac{\sqrt{p^2 -4(u^2 -m^2)}}{2|u|} \right).
\end{eqnarray}

\vspace{5mm}
\newpage
\vspace{5mm}
\newpage

\begin{table}
\begin{tabular}{ddddd}
$|\kappa|$ & $\langle\bar{\psi}\psi\rangle$ & 
$|\langle\psi^\dagger \psi\rangle|$ & $m_d$ & $|\mu_d|$ \\
\hline
0               & $-N/2\pi l^2$ & 0                & 
$\omega_s /l$  & 0 \\
$\alpha$        & $-N/4\pi l^2$ & $N/4\pi l^2$ & 
$\omega_s /2l$ & $\omega_s /2l$ \\
$2\alpha$       & 0             & $N/2\pi l^2$ & 
0               & $\omega_s /l$ 
\end{tabular}\vspace{12pt}
\caption{Condensates and solutions to the Schwinger-Dyson equation 
for each $\kappa$.\label{tab:config}}
\end{table}


\begin{references}
\bibitem[*]{taichi} Email address: taichi@newton.skku.ac.kr
\bibitem[\dagger]{kato} Email address: hiro@eken.phys.nagoya-u.ac.jp
\bibitem{frad}

 F. Wilczek,
 Fractional Statistics and Anyon Superconductivity
 (World Scientific, Singapore, 1990);
 E. Fradkin, Field Theories of Condensed Matter Systems
 (Addison-Wesley, Reading, MA, 1991).

\bibitem{dorey}

 N. Dorey and N.E. Mavromatos, Nucl. Phys. B 386 (1992) 614;
 I.J.R. Aitchison and N. Mavromatos, Phys. Rev. B 53 (1996) 9321.

\bibitem{RVB}

 G. Baskaran and P.W. Anderson, Phys. Rev. B 37 (1988) 580;
 L. Ioffe and A. Larkin, Phys. Rev. B 39 (1989) 8988;
 N. Nagaosa and P.A. Lee, Phys. Rev. Lett. 64 (1990) 2450;
 P.A. Lee and N. Nagaosa, Phys. Rev B 46 (1992) 5621.
 
\bibitem{polch}

 R. Shankar, Physica A 77 (1991) 530;
 Rev. Mod. Phys. 66 (1994) 129;
 J. Polchinski, Nucl. Phys. B 422 (1994) 617;
 C. Nayak and F. Wilczek, Nucl. Phys. B 417 (1994) 359.

\bibitem{jain}

 J. Jain, Phys. Rev. Lett. 63 (1989) 199;
 Phys. Rev. B 40 (1989) 8079.

\bibitem{zhang}

 S.C. Zhang, T.H. Hansson and S. Kivelson,
 Phys. Rev. Lett. 62 (1989) 82;
 A. Lopez and E. Fradkin, Phys. Rev. B 44 (1991) 5246;
 B.I. Halperin, P.A. Lee and N. Read, Phys. Rev. B 47 (1993) 7312.

\bibitem{deser}

 S. Deser, R. Jackiw and S. Templeton,
 Phys. Rev. Lett. 48 (1982) 975;
 Ann. Phys. (N.Y.) 140 (1982) 372.

\bibitem{hosotani}

 Y. Hosotani, Phys. Lett. B 319 (1993) 332;
 Phys. Rev. D 51 (1995) 2022;
 D. Wesolowski and Y. Hosotani, Phys. Lett. B 354 (1995) 396.

\bibitem{IS} 

 T. Itoh and T. Sato, Phys. Lett. B 367 (1996) 290.

\bibitem{gusynin}

 V.P. Gusynin, V.A. Miransky and I.A. Shovkovy,
 Phys. Rev. Lett. 73 (1994) 3499; Phys. Rev. D 52 (1995) 4718.

\bibitem{klimenko}

 K.G. Klimenko, Z. Phys. C 54 (1992) 323;
 K.G. Klimenko, B.V. Magnitsky and A.S. Vshivtsev,
 Nuovo Cim. A 107 (1994) 439.

\bibitem{gusynin2}

 V.P. Gusynin, V.A. Miransky and I.A. Shovkovy,
 Phys. Lett. B 349 (1995) 477; Nucl. Phys. B 462 (1996) 249.

\bibitem{FKM}

 K. Farakos, G. Koutsoumbas and N.E. Mavromatos,
 Phys. Lett. B 431 (1998) 147.

\bibitem{semenoff}

 G.W. Semenoff, I.A. Shovkovy and L.C.R. Wijewardhana,
 Mod. Phys. Lett. A 13 (1998) 1143;
 K. Farakos and N.E. Mavromatos, Athens/Oxford Preprint NTUA 67/97,
 OUTP-97-58P (cond-mat/9710288);
 N.E. Mavromatos and A. Momen, Mod. Phys. Lett. A 13 (1998) 1765.

\bibitem{appel}

 T. Appelquist, M. Bowick, D. Karabali and L.C.R. Wijewardhana,
 Phys. Rev. D 33 (1986) 3704;
 T. Appelquist, D. Nash and L.C.R. Wijewardhana,
 Phys. Rev. Lett. 60 (1988) 2575.

\bibitem{NN}

 D. Nash, Phys. Rev. Lett. 62 (1989) 3024;
 H. Nakatani, Comment at 1988 International Workshop on New Trends
 in Strong Coupling Gauge Theories, Aug. 24-27, 1988, Nagoya
 (unpublished, not included in the Proceedings);
 T. Appelquist and D. Nash, Phys. Rev. Lett. 64 (1990) 721.

\bibitem{shpagin}

 A.V. Shpagin, Dynamical mass generation in (2+1)
 dimensional electrodynamics in an external magnetic field
 (hep-ph/9611412).

\bibitem{IK}

 T. Itoh and H. Kato, Phys. Rev. Lett. 81 (1998) 30.

\bibitem{schwinger}

 J. Schwinger, Phys. Rev. 82 (1951) 664;
 C. Itzykson and J.B. Zuber, Quantum Field Theory
 (McGraw-Hill, New York, 1980) p. 100.

\bibitem{corn}

 J. Cornwall, R. Jackiw and E. Tomboulis,
 Phys. Rev. D 10 (1974) 2428.

\bibitem{redlich}

 A.N. Redlich, Phys. Rev. Lett. 52 (1984) 18;
 Phys. Rev. D 29 (1984) 2366;
 K. Ishikawa, Phys. Rev. Lett. 53 (1984) 1615;
 Phys. Rev. D 31 (1985) 1432.

\bibitem{vafa}

 C. Vafa and E. Witten,
 Comm. Math. Phys. 95 (1984) 257; Nucl. Phys. B 234 (1984) 173.

\bibitem{dittrich}

 W. Dittrich and M. Reuter, Effective Lagrangian in
 Quantum Electrodynamics (Springer-Verlag, Berlin, 1985).

\bibitem{chodos}

 A. Chodos, K. Everding and D.A. Owen, Phys. Rev. D 42 (1990) 2881.

\bibitem{jackiw}

 R. Jackiw, Phys. Rev. D 9 (1974) 1686.

\bibitem{lykken}

 J.D. Lykken, J. Sonnenschein and N. Weiss,
 Phys. Rev. D 42 (1990) 2161;
 Int. J. Mod. Phys. A 6 (1991) 1335; A 6 (1991) 5155.

\bibitem{zeitlin}

 V. Zeitlin, Phys. Lett. B 352 (1995) 422.

\bibitem{hosotani2}

 Y. Hosotani, Phys. Rev. Lett. 62 (1989) 2785.

\bibitem{NS}

 A.J. Niemi and G.W. Semenoff, Phys. Rev. Lett. 51 (1983) 2077.

\bibitem{hong}

 D.K. Hong and S.H. Park, Phys. Rev. D 47 (1993) 3651;
 K.-I. Kondo, T. Ebihara, T. Iizuka, E. Tanaka,
 Nucl. Phys. B 434 (1995) 85.

\bibitem{kondo}

 K.-I. Kondo and P. Maris,
 Phys. Rev. Lett. 74 (1995) 18; Phys. Rev. D 52 (1995) 1212;
 D.K. Hong, Phys. Rev. D 57 (1998) 1313.

\bibitem{coleman}

 S. Coleman and B. Hill, Phys. Lett. B 159 (1985) 184.

\bibitem{SSW}

 G.W. Semenoff, P. Sodano and Y.-S. Wu,
 Phys. Rev. Lett. 62 (1989) 715.

\bibitem{zeitlin2}

 V. Zeitlin, Mod. Phys. Lett. A 12 (1997) 877.

\bibitem{GR}

 I.S. Gradshteyn and I.M. Ryzhik,
 Table of Integrals, Series and Products
 (Academic Press, San Diego, 1994).

\bibitem{AH}

 J.O. Andersen and T. Haugset, Phys. Rev. D 51 (1995) 3073;
 P. Cea, Phys. Rev. D 55 (1997) 7985;
 P. Cea and L. Tedesco, Phys. Lett. B 425 (1998) 345.

\bibitem{kanemura}

 S. Kanemura and T. Matsushita, Phys. Rev. D 56 (1997) 1034.

\bibitem{das}

 A. Das and M. Hott, Phys. Rev. D 53 (1996) 2252.

\bibitem{ichinose}

 I. Ichinose and M. Onoda, Nucl. Phys. B 435 (1995) 637.

\bibitem{DY}

 S. Deser and Z. Yang, Mod. Phys. Lett. A 4 (1989) 2123;
 J. Hong, Y. Kim and P.Y. Pac, Phys. Rev. Lett. 64 (1990) 2230;
 R. Jackiw and E.J. Weinberg, Phys. Rev. Lett. 64 (1990) 2234.

\bibitem{IKSY}

 T. Itoh, Y. Kim, M. Sugiura and K. Yamawaki,
 Prog. Theor. Phys. 93 (1995) 417;
 K.-I. Kondo, Nucl. Phys. B 450 (1995) 251;
 M. Sugiura, Prog. Theor. Phys. 97 (1997) 311.

\end{references}
\end{document}